\documentclass[epsf,12pt]{article}   
\input epsf
\newlength{\dinwidth} 
\newlength{\dinmargin}
\begin{document}
\newcommand{\GeVs}      {\mbox{${\rm GeV}^2$}}
\newcommand{\GeV}       {\mbox{${\rm GeV}$}}
\newcommand{\MeV}       {\mbox{${\rm MeV}$}}
\newcommand{\qsq}       {\mbox{${Q^2}$}}
\newcommand{\rhod}      {\mbox{${\rho^0}$}}
\newcommand{\wgp}       {\mbox{${W}$}}
\newcommand{\empz}      {\mbox{${\delta}$}}
\newcommand{\phid}      {\mbox{${\phi}$}}
\newcommand{\jpsid}     {\mbox{${J/\psi}$}}
\newcommand{\yjb}       {\mbox{${y_{_{JB}}}$}}
\newcommand{\ys}        {\mbox{$y_{\Sigma}$}}
\newcommand{\xes}       {\mbox{$x_{e\Sigma}$}}
\newcommand{\wes}       {\mbox{$W_{e\Sigma}$}}
\newcommand{\eda}       {\mbox{$E_{DA}^{\prime}$}}
\newcommand{\ra}        {\mbox{$\rightarrow$}}
\newcommand{\gsp}       {\mbox{$\gamma^* p$}}
\newcommand{\xda}       {\mbox{$x_{DA}$}}
\newcommand{\qda}       {\mbox{$Q^{2}_{DA}$}}
\newcommand{\qe}        {\mbox{$Q^{2}_{e}$}}
\newcommand{\gsubh}     {\mbox{$\gamma_{_{H}}$}}
\newcommand{\te}        {\mbox{$\theta_{e}^{\prime}$}}
\newcommand{\Ee}        {\mbox{$E_{e}^{\prime}$}}
\newcommand{\gh}        {\mbox{$\gamma_{h}$}}
\newcommand{\mrsdm}     {\mbox{MRSD$_-^{\prime}$\ }}
\newcommand{\mrsdz}     {\mbox{MRSD$_0^{\prime}$\ }}
\newcommand{\ftwo}      {\mbox{$F_{2}$}}
\newcommand{\ftwoccb}   {\mbox{$F_{2}^{c\bar c}$}}
\newcommand{\ptr}       {\mbox{$p_{T}$}}
\newcommand{\ccb}       {\mbox{$c\bar{c}$}}
\newcommand{\dspm}      {\mbox{$D^{\ast \pm}$}}
\newcommand{\ds}        {\mbox{$D^{\ast}$}}
\newcommand{\dz}        {\mbox{$D^{0}$}}
\newcommand{\mc}        {\mbox{$m_{c}$}}
\newcommand{\dm}        {\mbox{$\Delta M$}}
\newcommand{\dsp}       {\mbox{$D^{\ast +}$}}
\newcommand{\xd}        {\mbox{$x(D^{\ast})$}}
\newcommand{\ptds}      {\mbox{$p_T(D^{\ast})$}}
\newcommand{\etads}     {\mbox{$\eta(D^{\ast})$}}
\newcommand{\mkpi}      {\mbox{$M_{K\pi}$}}
\newcommand{\mdz}       {\mbox{$M(D^0)$}}
\newcommand{\mkpipis}   {\mbox{$M_{K\pi\pi_{s}}$}}
\newcommand{\ps}        {\mbox{$\pi_{s}$}}
\newcommand{\sleq} {\raisebox{-.6ex}{${\textstyle\stackrel{<}{\sim}}$}}
\newcommand{\sgeq} {\raisebox{-.6ex}{${\textstyle\stackrel{>}{\sim}}$}}
\renewcommand{\thefootnote}{\arabic{footnote}}
\begin{titlepage}
\vspace{5.5 cm}

\title {
        \bf\LARGE  Measurement of $D^{\ast \pm}$ production \\
                   and the charm contribution to $F_2$\\
                   in deep inelastic scattering at HERA }

\author{ZEUS Collaboration \\
}
\date{ }
\maketitle
\vspace{3 cm}
\begin{abstract}
\par
The production of \dspm(2010) mesons 
in deep inelastic scattering has been measured in
the ZEUS detector at HERA using an integrated luminosity
of 37 pb$^{-1}$. The decay channels 
$D^{\ast +}\rightarrow D^0 \pi^+ $ (+\,c.c.), with 
$D^0 \rightarrow K^- \pi^+$ or $D^0 \rightarrow K^- \pi^- \pi^+ \pi^+$,
have been used to identify the $D$ mesons. 
The $e^+p$ cross section for inclusive \dspm\ production
with $1<Q^2<600\,$GeV$^2$ and $0.02<y<0.7$ is 
$8.31 \pm 0.31 \mbox{(stat.)}^{+0.30}_{-0.50}\mbox{(syst.)}$ nb
in the kinematic region {$1.5<\ptr(\dspm)<15$\,GeV and
$\left|\,\eta(\dspm)\,\right|<1.5$}.
Differential cross sections are consistent with a
next-to-leading-order  perturbative-QCD calculation when using 
charm-fragmentation models which take into account
the interaction of the charm quark with the proton remnant.
 The observed cross section is extrapolated to the full
kinematic region in \ptr(\dspm) and $\eta$(\dspm) in order to determine 
the charm contribution, \ftwoccb$(x,Q^2)$, to the proton
structure function. The ratio \ftwoccb/\ftwo\ rises from $\simeq$10\% at 
$Q^2  \simeq$ 1.8 GeV$^2$ 
to $\simeq$30\% at $Q^2 \simeq $130 GeV$^2$ for $x$ values in the range $10^{-4}$
to $10^{-3}$.

\end{abstract}

\vspace{-17.5 cm}
{\noindent
Eur. Phys. J. {\bf C}12 (2000) 35-52}
\setcounter{page}{0}
\thispagestyle{empty}
\eject
\end{titlepage}

\newpage
{
\parindent0.cm                                                                                     
\parskip0.3cm plus0.05cm minus0.05cm                                                               
\def\3{\ss}                                                                                        
\newcommand{\address}{ }                                                                           
\newcommand{\author}{ }                                                                          
                                                   %
\begin{center}                                                                                     
{                      \Large  The ZEUS Collaboration              }                               
\end{center}                                                                                       
  J.~Breitweg,                                                                                     
  S.~Chekanov,                                                                                     
  M.~Derrick,                                                                                      
  D.~Krakauer,                                                                                     
  S.~Magill,                                                                                       
  B.~Musgrave,                                                                                     
  A.~Pellegrino,                                                                                   
  J.~Repond,                                                                                       
  R.~Stanek,                                                                                       
  R.~Yoshida\\                                                                                     
 {\it Argonne National Laboratory, Argonne, IL, USA}~$^{p}$
% ...
\newpage                                                                                           
\noindent                                                           %
$^{\    1}$ now visiting scientist at DESY \\                                                      
$^{\    2}$ also at IROE Florence, Italy \\                                                        
% ...
$^{  35}$ partially supported by the Foundation for German-Russian Collaboration                   
DFG-RFBR \\ \hspace*{3.5mm} (grant no. 436 RUS 113/248/3 and no. 436 RUS 113/248/2)\\              
% ...
\newpage   
                                                           %
                                                           %
\begin{tabular}[h]{rp{14cm}}                                                                       
$^{a}$ &  supported by the Natural Sciences and Engineering Research                               
          Council of Canada (NSERC)  \\                                                            
$^{b}$ &  supported by the FCAR of Qu\'ebec, Canada  \\                                            
$^{c}$ &  supported by the German Federal Ministry for Education and                               
          Science, Research and Technology (BMBF), under contract                                  
          numbers 057BN19P, 057FR19P, 057HH19P, 057HH29P, 057SI75I \\                              
$^{d}$ &  supported by the MINERVA Gesellschaft f\"ur Forschung GmbH, the                          
German Israeli Foundation, and by the Israel Ministry of Science \\                                
$^{e}$ &  supported by the German-Israeli Foundation, the Israel Science                           
          Foundation, the U.S.-Israel Binational Science Foundation, and by                        
          the Israel Ministry of Science \\                                                        
$^{f}$ &  supported by the Italian National Institute for Nuclear Physics                          
          (INFN) \\                                                                                
$^{g}$ &  supported by the Japanese Ministry of Education, Science and                             
          Culture (the Monbusho) and its grants for Scientific Research \\                         
$^{h}$ &  supported by the Korean Ministry of Education and Korea Science                          
          and Engineering Foundation  \\                                                           
$^{i}$ &  supported by the Netherlands Foundation for Research on                                  
          Matter (FOM) \\                                                                          
$^{j}$ &  supported by the Polish State Committee for Scientific Research,                         
          grant No. 115/E-343/SPUB/P03/154/98, 2P03B03216, 2P03B04616,                             
          2P03B10412, 2P03B03517, and by the German Federal                                        
          Ministry of Education and Science, Research and Technology (BMBF) \\                     
$^{k}$ &  supported by the Polish State Committee for Scientific                                   
          Research (grant No. 2P03B08614 and 2P03B06116) \\                                        
$^{l}$ &  partially supported by the German Federal Ministry for                                   
          Education and Science, Research and Technology (BMBF)  \\                                
$^{m}$ &  supported by the Fund for Fundamental Research of Russian Ministry                       
          for Science and Edu\-cation and by the German Federal Ministry for                       
          Education and Science, Research and Technology (BMBF) \\                                 
$^{n}$ &  supported by the Spanish Ministry of Education                                           
          and Science through funds provided by CICYT \\                                           
$^{o}$ &  supported by the Particle Physics and                                                    
          Astronomy Research Council \\                                                            
$^{p}$ &  supported by the US Department of Energy \\                                              
$^{q}$ &  supported by the US National Science Foundation                                          
\end{tabular}                                                                                      
                                                           %
                                                           %
}

\newpage


\section{\bf Introduction}
\label{s:intro}
The first HERA measurements of the charm contribution, \ftwoccb, to the proton 
structure function \ftwo\ were reported by the H1 and ZEUS collaborations 
from the analyses of their 1994 deep inelastic scattering (DIS) data 
sets \cite{h194,z94}. 
These early results, which were statistically limited,
revealed a steep rise of \ftwoccb\ as Bjorken-$x$ decreases.
At the lowest accessible $x$ values, it was found that 
around 25\% of DIS events contained open charm, 
in contrast to the EMC fixed target measurements~\cite{fixtgt}
in the high-$x$ region where the charm contribution is small.
Given the large charm content, the correct theoretical treatment of charm 
for \ftwo\ analyses in 
the HERA regime has become essential. More detailed
measurements of charm production will aid such analyses.

\begin{figure}[hbtp]
\epsfysize=10cm
\centerline{\epsffile{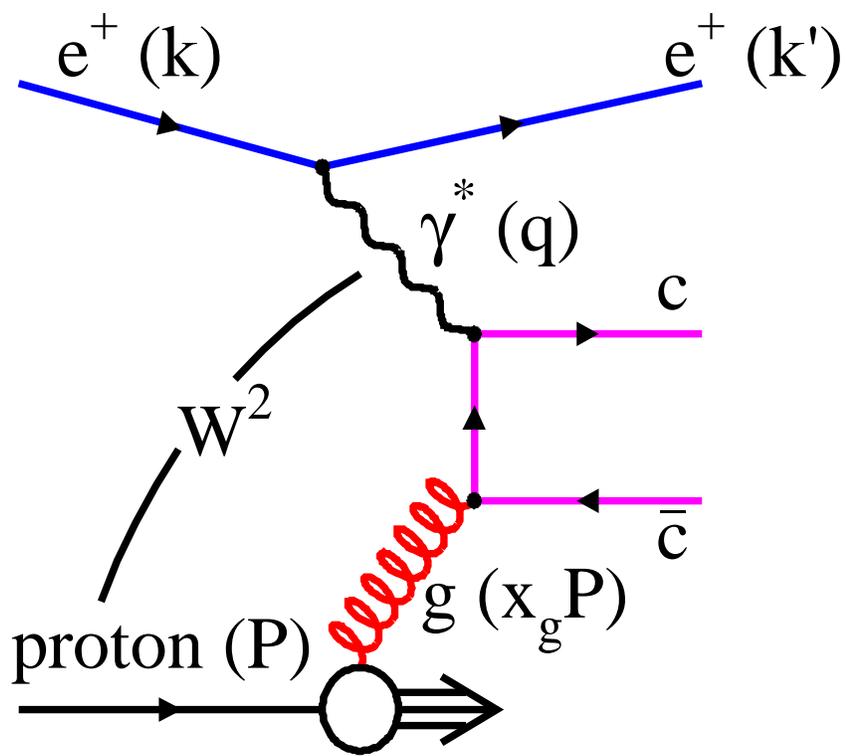}}
\caption{\it Diagram of the boson-gluon-fusion process 
in $e^+p$ collisions.}
\label{f:bgf}
\end{figure}

The early 
results~\cite{h194,z94} suggested that the production dynamics of charmed
mesons in $ep$ collisions are
dominated by the boson-gluon-fusion (BGF) mechanism shown in 
Fig.~\ref{f:bgf}. In this case, the reactions $e^+p \rightarrow e^+ D^{\ast \pm}
 X$ are sensitive to the gluon distribution 
in the proton~\cite{h195}. The measurement of charm production can also provide 
tests of 
perturbative QCD (pQCD), in particular, tests of the hard 
scattering factorization theorem, which states that the same, universal,
 gluon distribution
should contribute to both \ftwo~ and \ftwoccb. In addition, 
the presence of two large scales, namely,
the virtuality of the
exchanged boson ($Q^2$) and  the square of the charm-quark mass
($m_c^2$), provides a
testing ground for  resummation techniques.

This paper reports  a measurement
of \dspm(2010) production using the 1996 and 1997 data sets collected with the ZEUS 
detector, corresponding to an integrated luminosity of 
37 pb$^{-1}$. During this period, HERA collided $E_e$ = 27.5 GeV positrons with
$E_p$ = 820 GeV protons, yielding a center-of-mass energy, $\sqrt{s}$, of
300 GeV. This larger data sample, together with 
some improvements to the ZEUS detector,
allows an extension of the kinematic range to both larger and smaller \qsq. 
The \dspm\ production is investigated in the decay channel
$\dsp(2010)\rightarrow \dz(1864) 
\pi^+_s$ ($+$ c.c.), where $\pi_s^+$ refers to a slow 
$\pi^+$~\cite{slowpi},
followed by the $D^0$  decay channels $D^0 \rightarrow  
K^-\pi^+$ ($+$ c.c.) 
or $D^0 \rightarrow K^-\pi^+\pi^-\pi^+$ ($+$ c.c.). These two
final states will be referred to as the $K2\pi$ and $K4\pi$ channels,
respectively.

Following a brief discussion of three pQCD calculations
for \dspm\ production in DIS, a description of the experiment and details
of the data
analysis are provided. Next, the cross sections for \dspm\ are
presented and are compared to the QCD predictions. 
A possible cause of the observed discrepancies between the data and 
the QCD results is discussed. Finally, the \dspm\
results are used to obtain the charm contribution to \ftwo.

\section{\bf Charm quark production models}
\label{s:models}

General agreement was found between the earlier HERA $D^{\ast \pm}$ data and 
the results from a Monte Carlo program HVQDIS~\cite{hvqdis} for
the production of
heavy quarks and their subsequent fragmentation to hadrons.
This program is based on next-to-leading-order (NLO) calculations~\cite{smith} 
of order $\alpha_s^2$ in the coefficient functions
in the so-called fixed-flavor-number scheme (FFNS). In this scheme,
the number of active quark flavors is fixed, independent of $Q^2$. 
Only light quarks ($u,~d,~s$) are included in the initial-state 
proton and charm quarks are produced exclusively
by BGF with NLO processes~\cite{smithlaenenneerven}. 
The presence of the two large scales, $Q^2$ and $m_c^2$,
can spoil the convergence of the perturbative
series because the neglected terms of orders higher than $\alpha_s^2$
contain log$(Q^2/m_c^2)$ factors
that can become large. Therefore the results of HVQDIS are expected
to be most accurate at $Q^2 \approx m_c^2$ and to become 
less reliable when $Q^2 \gg m_c^2$.

In this high $Q^2$ region, the $c$  quark can be treated as massless, as
implemented in the zero-mass, variable-flavor-number scheme (ZM-VFNS).
In this scheme, 
the resummation of large logarithms of $\qsq/ m_c^2$~\cite{acot,buza} 
results in  
a charm density which is added as a fourth flavor and which is then evolved 
in the same way as the light quark densities.
At intermediate \qsq\ values, the two schemes need to be merged.
One way in which this is done
is described by ACOT~\cite{acot} and by Collins~\cite{collins}. 
An alternative matching method  has been proposed by MRST~\cite{mrst}.

A third method for  modelling \dspm\ production has recently
been suggested by BKL~\cite{bkl}. This 
tree-level pQCD calculation, applicable for $p_T \ge m_c$, considers 
the hadronization of the $(c\bar{q})$-state
into a \ds, in contrast to hadronizing an isolated $c$-quark. 
 The \ds\ is created from both color singlet and color octet
configurations of the light and heavy quarks. 
Results from $e^+e^-$ annihilation imply that the octet 
contribution is small. However,
 the singlet contribution alone underestimates~\cite{bkl}
the ZEUS data on the photoproduction of charm~\cite{zeus_ds_php}.
This calculation has been extended to DIS charm production and is
compared to the \dspm\ data  reported here.

\section{\bf Experimental setup}
\label{s:ZEUS}

ZEUS is a multipurpose detector which
has been described in detail elsewhere \cite{zeusdetector}. 
The key component for this analysis is the central tracking detector
(CTD) which operates in a magnetic field of 1.43 T provided by a
thin superconducting solenoid. The CTD~\cite{CTD} is a drift chamber
consisting of 72 cylindrical layers,
arranged in 9 superlayers covering the polar angle\footnote{
The ZEUS coordinate system is defined as right-handed 
with the $Z$ axis pointing in the forward (proton beam) 
direction and the $X$ axis 
horizontal pointing towards the center of HERA. The origin is at the nominal 
$ep$ interaction point. The polar angle $\theta$ 
is defined with respect to the positive $Z$ direction.}
region $15^\circ < \theta < 164^\circ$.
The transverse momentum resolution for full-length tracks is 
$\sigma(p_{\rm T}) / p_{\rm T} = 
 0.0058\, p_{\rm T} \bigoplus 0.0065 \bigoplus 0.0014/p_{\rm T}$ 
($p_{\rm T}$ in GeV). The CTD
was also used to establish an interaction vertex for each event.

The uranium-scintillator sampling calorimeter (CAL) surrounds the solenoid. 
The CAL is hermetic and consists of 5918 cells, each read
out by two photomultiplier tubes. The CAL contains three
parts, the forward (FCAL), barrel (BCAL) and rear (RCAL),
with longitudinal segmentation into electromagnetic and hadronic sections.
The energy resolutions, as measured in test beams, are
$\sigma /E$ = 0.18/$\sqrt{E(\mbox{GeV})}$ and 
0.35/$\sqrt{E(\mbox{GeV})}$ for electrons and hadrons, 
respectively~\cite{CALRes}.

The position of positrons scattered close to the positron beam 
direction is determined by a scintillator strip detector (SRTD)
\cite{SFTVTX}. The luminosity was measured from the rate of the 
bremsstrahlung process, $ e^+ p \rightarrow
e^+ \gamma p$, where the photon is measured by
a lead/scintillator calorimeter\,\cite{LUMI}
located at $Z\,=\,-\,107$\,m in the HERA tunnel. 

\section{Kinematics and reconstruction of variables}
The reaction $e^+(k) + p (P)  \rightarrow e^+(k^{\prime}) + X$ at fixed 
squared center-of-mass energy, $s = (k+P)^2$, is described in terms of \qsq\ = $-q^2 = 
-(k-k^{\prime})^2$ and  Bjorken-$x=Q^2/(2P\cdot q)$. 
The fractional energy transferred to the proton in its rest frame
is $y = Q^2/(sx)$. The virtual photon ($\gamma^{\ast}$)-proton center-of-mass
energy $W$, given by $W^2 = (q+P)^2$, is also used, see Fig.~\ref{f:bgf}.

In neutral current (NC) $e^+p$ DIS, both the final-state positron, 
with energy \Ee~ and angle \te, 
and the hadronic system (with a characteristic angle \gh, which, in the simple
quark-parton model, is the polar angle of the struck quark) can be measured.
The scattered positron was identified using an algorithm based on a neural
network~\cite{sinistra}. 
CAL cells were combined to form clusters and combinations of
these clusters and CTD tracks were used to reconstruct energy-flow objects 
(EFO's)~\cite{zufos,gennady}. For perfect detector 
resolution and acceptance, the quantity $\delta \equiv \Sigma_i(E_i-p_{z,i})$ 
is equal to 2$E_e$ (55 GeV).
Here, $E_i$ and $p_{z,i}$ are the energy and longitudinal component of the
momentum assigned to the $i$-th EFO. The sum runs over all 
EFO's including those assigned to the scattered positron. 

In the $K2\pi$ analysis, \qsq\ was reconstructed from the scattered positron 
(\qe)  with the electron method~\cite{breit}
and $y$ with the so-called $\Sigma$ method~\cite{sigmamethod}
\begin{equation}
\ys = \frac{\delta_{had}}{\delta},
\end{equation}
where $\delta_{had}$ is calculated in the same way as $\delta$ 
but excluding the EFO's assigned to the scattered positron.
Bjorken-$x$ (\xes) and $W$ (\wes)  are then defined by a combination of 
\qe\ and \ys\ ($e\Sigma$ method). 

The fractional momentum of the \dspm\ in the $\gamma^* p$ system is defined as 
\begin{equation}
\xd = \frac{2 |\vec{p}^{~*}(\ds)|}{W},
\end{equation}
where $\vec{p}^{~*}(\ds)$ is the \dspm\ momentum in the $\gamma^* p$ 
center-of-mass frame.
For the boost to the 
$\gamma^* p$ system, the virtual-photon vector was reconstructed 
using the double-angle ($DA$) estimator~\cite{breit} of the scattered 
positron energy, \eda.
In this method, only the angles $\theta_e^{\prime}$ and $\gamma_h$
are used~\cite{zbreit}. 
\eda\ is less sensitive to radiative effects at the leptonic vertex than 
the scattered $e^+$ energy determined using the electron method.

For the $K4\pi$ analysis, \qsq, $x$ and $W$ were determined using the
$DA$ estimators ($Q^2_{DA}, x_{DA}$ and $W_{DA}$).

\section{Monte Carlo simulation}
\label{s:mc}
A GEANT 3.13-based~\cite{geant} Monte Carlo (MC) simulation program which 
incorporates the best current
knowledge of the ZEUS detector and trigger was used to
correct the data for detector and acceptance effects.
The event generator used 
for the simulation of the QED 
radiation from the leptonic vertex
was  RAPGAP~\cite{rapgap} 
interfaced to HERACLES 4.1~\cite{heracles}. 
The charm quarks were produced in the  
BGF process calculated at leading order (LO). 
The charm mass was set to 1.5~GeV.  
The GRV94HO~\cite{grv94}
parton distribution functions (pdf's) were used for the proton. 
Fragmentation was carried out using
the Lund 
model~\cite{lund}, as implemented in JETSET 7.4~\cite{jetset},
with the full parton shower option.
The fraction of the original $c$-quark momentum which is carried by the
\dspm\ is determined from the
`SLAC' fragmentation
function, which is equivalent to the
Peterson model~\cite{peterson}, with the fragmentation 
parameter $\epsilon$ set to 0.035~\cite{opal0.035}. 
The HERWIG 5.9~\cite{herwig} event generator was also used,
with the same pdf's and $c$-quark mass as used in RAPGAP,
to investigate the effects of fragmentation.

Generated events with at least one 
$\dsp\rightarrow \dz \pi^+_s \rightarrow  (K^-\pi^+$ or $K^-\pi^+\pi^-\pi^+) 
\pi^+_s$ (or c.c.) 
were selected. These events were then processed through the 
detector and trigger simulation and through the same reconstruction program as 
was used for the data.

\section{Event selection}
\label{s:sele}

\subsection{Trigger}
\label{ss:trigger}
Events were selected online with a three-level 
trigger~\cite{zeusdetector}.
At the first level (FLT), inclusive 
DIS events are  triggered by the presence of 
an isolated electromagnetic cluster in the RCAL or any energy deposition in
excess of 3~GeV in any electromagnetic section of the CAL~\cite{trigger}.
During high-luminosity periods, when the rate was high, a coincidence with
an FLT track was also required. Tracks at the FLT are defined as a series of 
CTD hits pointing to the nominal interaction point. The
efficiency of this trigger, with respect to the calorimeter-only trigger,
was greater than 99.5\% and in good agreement with the MC simulation.

At the second level, algorithms are applied to reduce the
non-$e^+p$ background. The full event information 
is available at the third level trigger (TLT). At this level,  events are 
accepted as DIS candidates if a high-energy scattered positron candidate 
is found within the CAL (`inclusive DIS trigger'). 
Because of the high rate of low-$Q^2$ events, 
this trigger was turned off in the region
around the RCAL beampipe during high-luminosity
operation. For the $K2\pi$ decay channel,
a $D^{\ast}$-finder (based on computing the $K\pi\pi$ mass 
using tracking information and selecting 
loosely around the \dspm\ mass) was available in the TLT.
Events at low \qsq\ were then kept by requiring a coincidence of an identified 
scattered positron anywhere in the CAL and a tagged \dspm\ candidate.
This will be referred to as the `\ds\ trigger'.

Using data selected from periods when both triggers were in use,
the relative efficiency of the  \ds\ trigger 
with respect to the inclusive DIS trigger 
is found to be about 80\% and independent of \qsq, $x$, \ptds\ 
and \etads\ within the measured 
kinematic regions.  The MC simulations  reproduce
this efficiency to an accuracy better than the statistical accuracy of the data
($\approx$ 2\%).

\subsection{Offline selection}
\label{ss:offline}
The DIS event selection was similar to that described in
an earlier publication~\cite{z95f2}; namely, the selection required:
\begin{itemize}
\item a positron, as identified by a neural network algorithm, with a corrected energy
above 10 GeV;
\item the impact point of the scattered positron on the RCAL was required to
lie outside the region 26$\times$14 cm$^2$ centered on the RCAL beamline;
\item 40 $< \delta <$ 65 GeV; and
\item a Z-vertex position $|Z_{vtx}|<$ 50 cm.
\end{itemize}
The DIS events were restricted to the kinematic region
\begin{itemize}
\item 1~$< Q^2 <600$~\GeVs\  and $0.02 < y < 0.7$.
\end{itemize}

\dspm\ candidates were reconstructed from CTD tracks
which were assigned to the reconstructed event vertex.
Only tracks with at least one hit in
the third superlayer of the CTD were considered.
This corresponds to an implicit requirement that  \ptr\ $>0.075$~GeV. 
Tracks were also required to have $| \eta | < 1.75$, where the pseudorapidity 
is defined as $\eta = -\ln (\tan \frac{\theta}{2})$.
The selected
tracks were in the region where the CTD performance is
well understood. For these tracks the reconstruction efficiency 
is above 95\%.

\dspm\ production was measured in the decay channel
$\dsp\rightarrow \dz \pi^+_s$ ($+$ c.c.),
which has a branching ratio of 0.683 $\pm$ 0.014~\cite{PDG},
followed by the $D^0$  decays $D^0 \rightarrow  
K^-\pi^+$ ($+$ c.c.) or $D^0 \rightarrow K^-\pi^+\pi^-\pi^+$ ($+$ c.c.).
The branching ratio for \dz\ to $K\pi$ ($K3\pi$) is 
0.0385 $\pm$ 0.0009 (0.076 $\pm$ 0.004)~\cite{PDG}. 
The remaining selection criteria were different for the two final states
and are discussed separately.
\subsection{Selection for the $K2\pi$ final state}

Pairs of oppositely-charged tracks were first combined to form a \dz\ 
candidate. Since
no particle identification was performed, the tracks were
alternatively assigned the masses of a charged kaon and a charged 
pion.  An additional slow track,
with charge opposite to that of the kaon track and   
assigned the pion mass (\ps),
was combined with the \dz\ candidate to form a \dspm\ candidate. 

The combinatorial background for the $K2\pi$ decay channel
was further reduced by requiring
\begin{itemize}
\item that the 
transverse momenta of the $K$ and the $\pi$ were greater than 
0.4~GeV, and that of the \ps\ was greater than 0.12~GeV. 
\end{itemize}
In addition, the momentum ratio requirement 
\begin{itemize}
\item
$p(\dz)/p(\ps) > 8.0$ 
\end{itemize}
was imposed. 
This requirement was used in the \ds\ trigger to reduce the 
rate of candidate events with large  
$\dm \equiv (\mkpipis - \mkpi)$, far from the signal region.

The \dspm\, kinematic region of the present
analysis was defined as
\begin{itemize}
\item 
 1.5 $<$ \ptr (\ds ) $<$ 15~GeV and $| \eta (\ds ) | < $ 1.5. 
\end{itemize}
Finally, the signal regions for the mass of the \dz\ candidate, $\mdz$, and 
$\dm $ were
\begin{itemize}
\item 1.80 $ < \mdz < 1.92 $~GeV, and
\item 143  $ < \dm  < 148  $~MeV.  
\end{itemize}

\subsection{Selection for the $K4\pi$ final state}

Permutations of two negatively- and two positively-charged tracks were 
first combined to form a \dz\ candidate. As for the $K2\pi$ channel,
the tracks were
alternatively assigned the masses of a charged kaon and a charged pion. 
An additional track,
with charge opposite to that of the kaon track and   
assigned the pion mass (\ps),
was combined with the \dz\ candidate to form a \dspm\ candidate. 
The combinatorial background for the $K4\pi$ decay channel
was reduced by requiring 
\begin{itemize}
\item
that the transverse momentum of the $K$ and each $\pi$ was greater than 
0.5 and 0.2~GeV, respectively, and that of the \ps\ was greater than 0.15~GeV.
\end{itemize} 
In addition, the momentum ratio requirement 
\begin{itemize}
\item $p(\dz)/p(\ps) > 9.5$ 
\end{itemize}
was imposed. 

The \dspm\ kinematic region was defined as
\begin{itemize}
\item  2.5 $<$ \ptr (\ds ) $<$ 15~GeV and $| \eta (\ds ) | < $ 1.5. 
\end{itemize}
The signal regions for $\mdz$ and 
$\dm \equiv (M_{K\pi\pi\pi\pi_{s}} - M_{K\pi\pi\pi})$ were
\begin{itemize}
\item 1.81 $ < \mdz < 1.91 $~GeV, and
\item 143  $ < \dm  < 148  $~MeV.  
\end{itemize}

\subsection{Mass distributions}

\begin{figure}[h]
\vspace{-6.5cm}
\epsfysize=18cm
\centerline{\epsffile{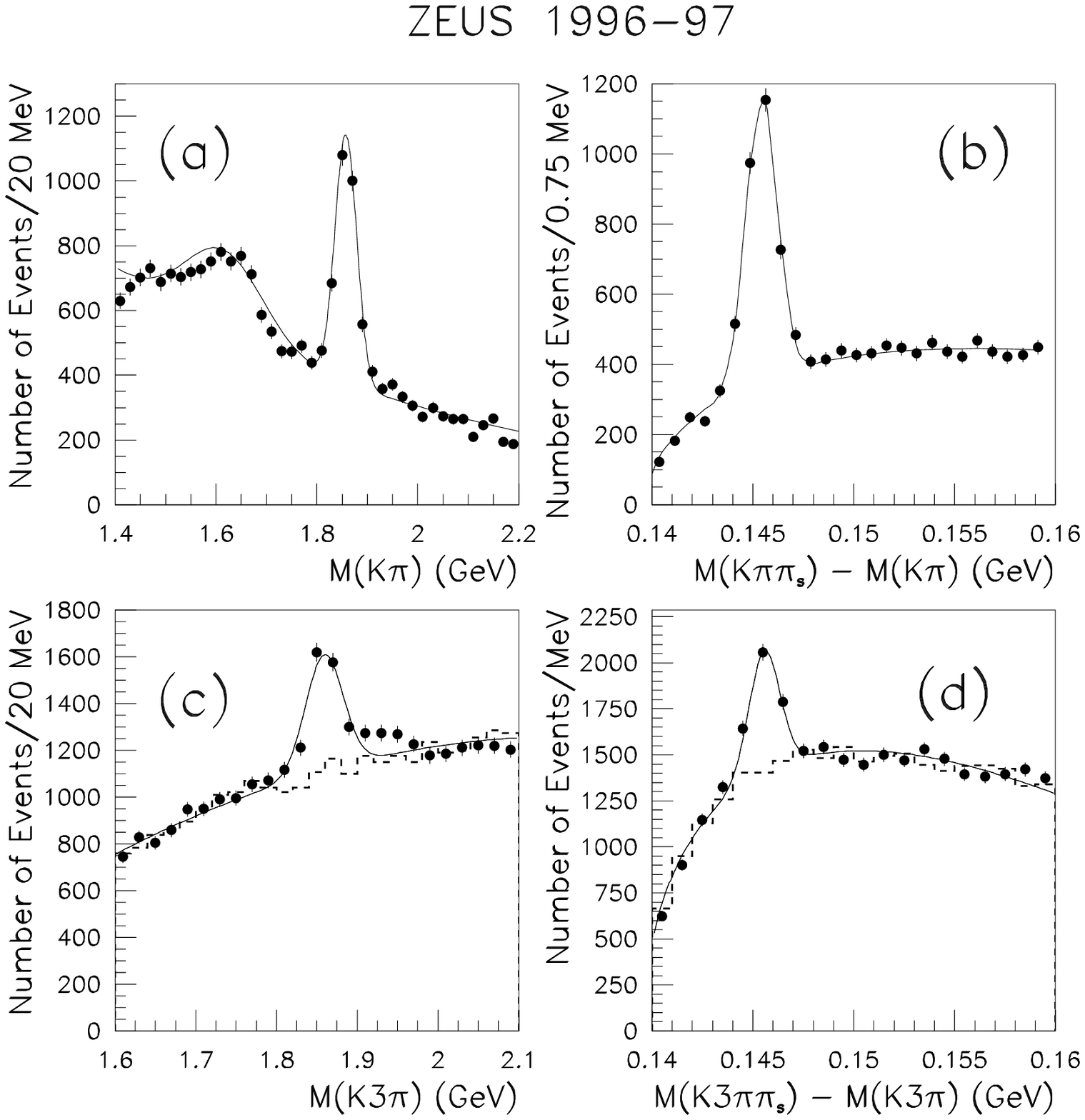}}
\caption{\it ZEUS data in the kinematic region $1<Q^2<600$ \GeVs
and $0.02 < y < 0.7$: (a) \mdz\ $= M(K\pi)$ and (b) \dm\ $= M(K\pi\pi_s)
-M(K\pi)$ for the $K2\pi$ final
state. The data are shown in the same kinematic region
for the $K4\pi$ final
state in (c) \mdz\ $= M(K3\pi)$ and (d) \dm\ $= M(K3\pi\pi_s) - M(K3\pi)$. 
The events in (a) and (c) are those for which 143 $<$ \dm $<$ 148 \MeV.
Similarly, (b) and (d) are shown for events in the \mdz\ signal region:
1.80-1.92 (1.81-1.91) \GeV for the $K2\pi$ ($K4\pi$) channel.
The solid curves are the results of the fits described in the text.
The dashed histograms in (c) and (d) represent the background distributions
obtained by restricting the sample to 
side-bands from the \dm\ and \mdz\ distributions, 
respectively.}
\label{f:signals}
\end{figure}

Figures~\ref{f:signals}(a) and (c) show the distributions of 
\mdz\  for 
candidates with \dm\ in the signal region for the two final states, while
Figs.~\ref{f:signals}(b) and (d) show the distributions of 
\dm\  for 
candidates with \mdz\ in the signal region.
Clear signals are observed around the expected mass values.

For the $K2\pi$ final state, a fit to the \mdz\ distribution of two Gaussians 
plus an exponentially falling background gives a peak at 
$\mdz=1863.2 \pm 0.8$~MeV and a width of $23 \pm 2$~MeV. 
The second Gaussian around 1.6~GeV in Fig.~\ref{f:signals}(a) 
originates primarily
from \dz\ decays to $K^-\pi^+\pi^0$ in which the neutral pion is not
reconstructed.
For the $K4\pi$ channel, the background level was determined
by using the side-bands outside the \dm\ signal region
(see the dashed histogram in Fig.~\ref{f:signals}(c)) to
 make a \mdz\ distribution.
The fit to the \mdz\ distribution, made by adding a Gaussian
to the background distribution, yielded 
$\mdz=1862.7 \pm 1.5$~MeV and a width of 20 $\pm$ 2~MeV. The deviations
of the data from the fit, visible in the region just above the signal
in Fig.~\ref{f:signals}(c),
are mostly due to the mass misassignments of the $K$ and $\pi$ candidates
with the same charge from \dz\ decay. This was verified by MC studies~\cite{Richard}. 
The mass values found for the \dz\ are consistent with 
the PDG~\cite{PDG} value of 1864.6$\pm$0.5~MeV.

The solid curve in Fig.~\ref{f:signals}(b) shows a binned maximum-likelihood 
fit to the 
\dm\ distribution from the $K2\pi$ channel using a
Gaussian plus a background of the form 
$A ( \dm - m_\pi)^B \exp [C(\dm - m_\pi)] $,
where $A$, $B$ and $C$ are free parameters and $m_\pi$ is the pion mass.
The fit to the $K2\pi$  plot gives
a peak at $\dm=145.44 \pm 0.05$~MeV, in good agreement 
with the PDG value 
of $145.397 \pm 0.030$~MeV, and a width of $0.79 \pm 0.05$ MeV,
in agreement with the experimental resolution. The multiplicative
exponential term is needed to describe the background suppression at large \dm,
which comes from the 
requirement on the momentum ratio $p(\dz)/p(\ps)$.

The solid curve in Fig.~\ref{f:signals}(d) shows a fit of the 
\dm\ distribution from the $K4\pi$ channel using a
Gaussian plus a background distribution obtained from the side-bands outside
the \mdz\ signal region (see the dashed histogram in Fig.~\ref{f:signals}(d)).
The fit yields
a peak at $\dm=145.61 \pm$ 0.05~MeV and a width of $0.78 \pm 0.07$ MeV.

%
For the $K2\pi$ channel, the number of \dspm\ events obtained from a fit to the
\dm\ distribution  in the restricted region of $Q^2$, $y$, \ptds\ and \etads\ is $2064 \pm 72$. The number of events in the $K4\pi$ channel is determined from the
\dm\ distribution using the side band method~\cite{zeus_ds_php}, which properly
accounts for the combinatorial background and the background arising from
the mass misassignments. The side bands, 
$1.74 < M(K\pi\pi\pi) < 1.79$~GeV and $1.93 < M(K\pi\pi\pi) < 1.98$~GeV, were normalized to the \dm\ distribution in the region $150 < \dm < 160$~MeV. The number of
\dspm\ events in the $K4\pi$ channel is found to be $1277 \pm 124$ in the restricted kinematic region of the measurement.

\section{Data characteristics}
\begin{figure}[hbtp]
\epsfysize=18cm
\centerline{\epsffile{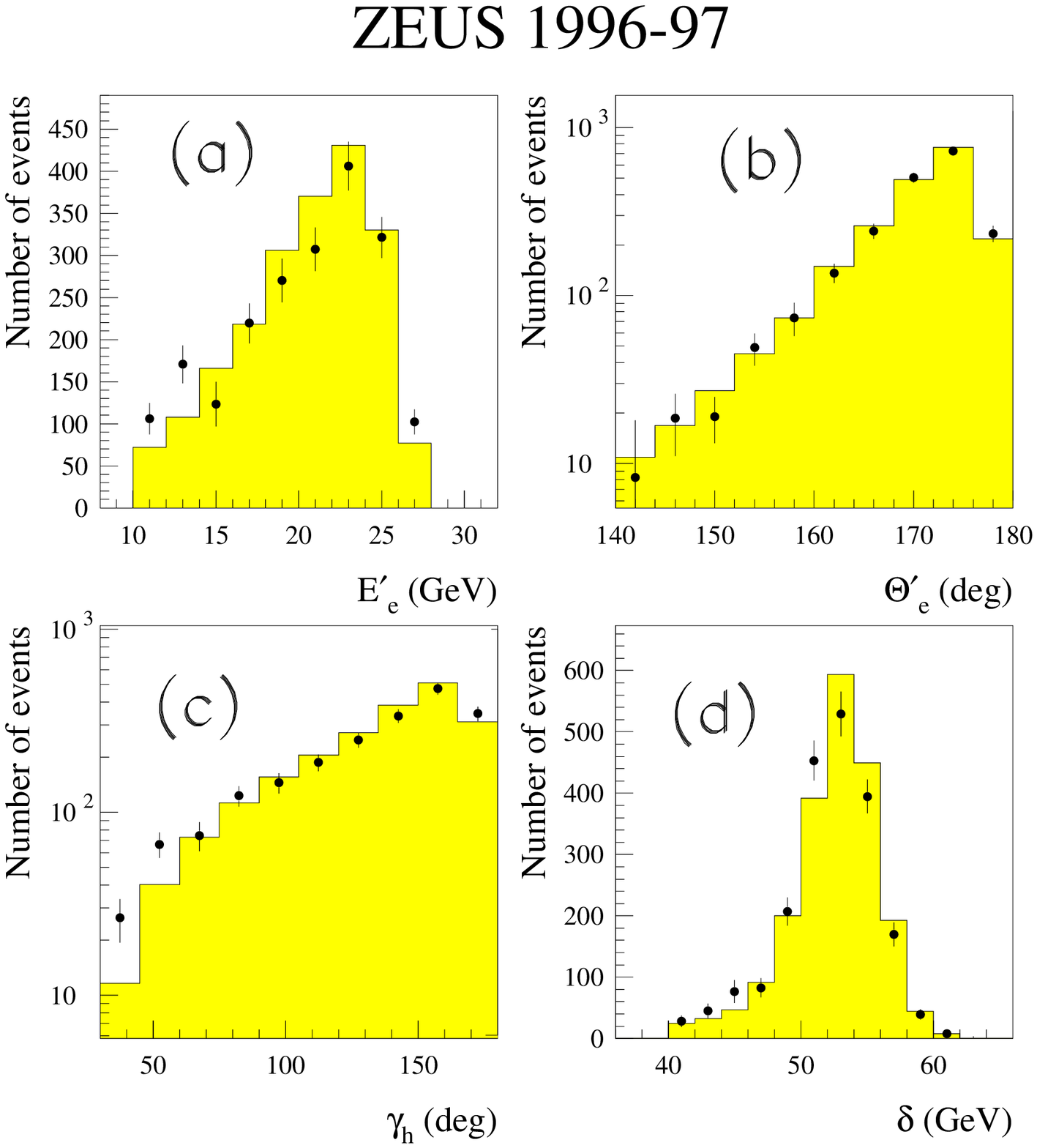}}
\caption{\it A comparison of reconstructed DIS quantities at 
the detector level for the $K2\pi$ data (points) and for the RAPGAP 
Monte Carlo simulation 
(shaded histogram): (a) the scattered positron energy, \Ee, (b) 
the scattered positron angle, \te, (c) the hadronic angle,
\gh\ and (d) \empz $\equiv  \Sigma_i(E_i - p_{z,i})$.}
\label{f:mcdatacontrol}
\end{figure}

\begin{figure}[hbtp]
\epsfysize=12cm
\centerline{\epsffile{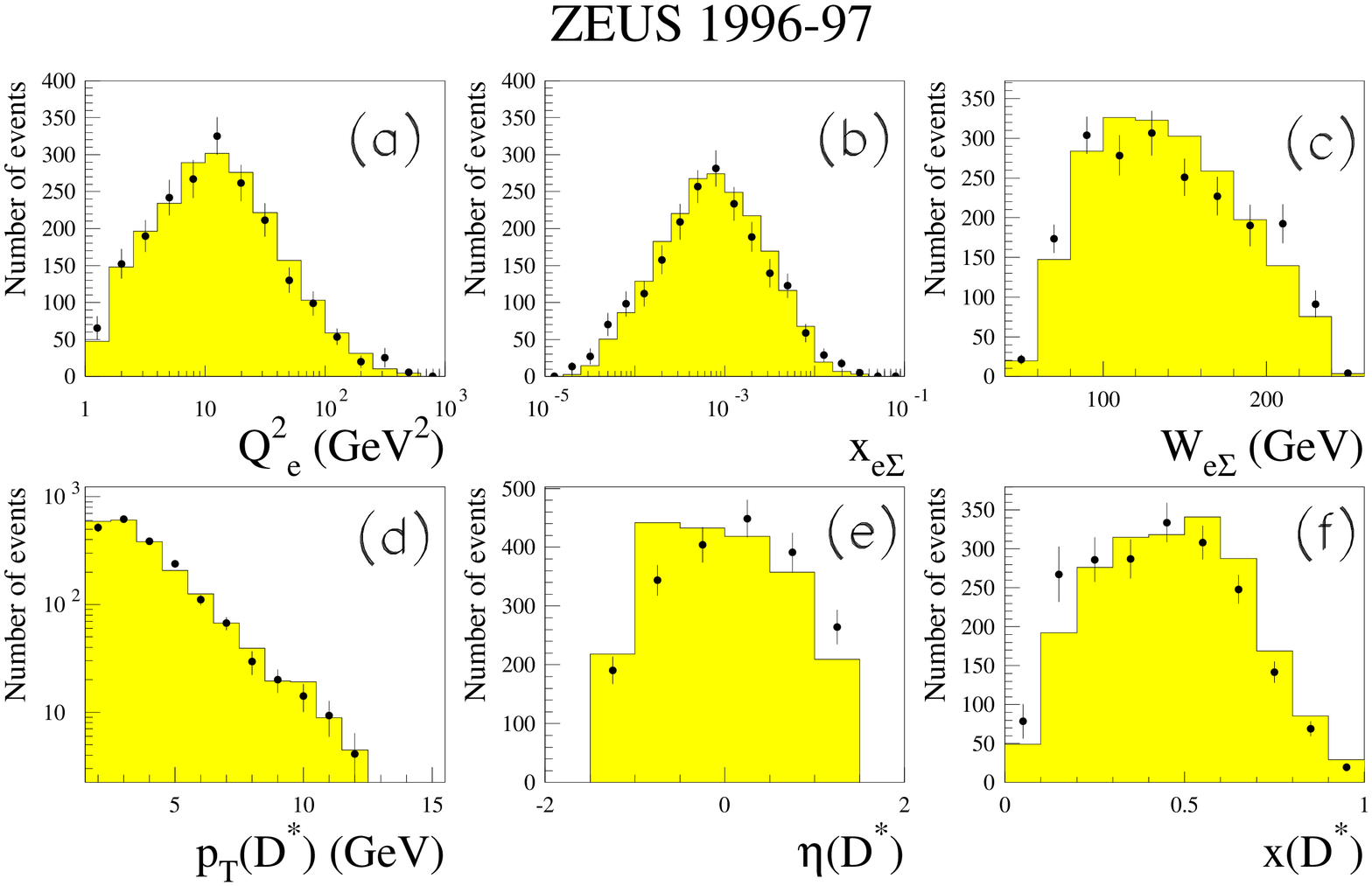}}
\caption{\it A comparison of the reconstructed DIS kinematic distributions at 
the detector level for the $K2\pi$ data (points) and for the RAPGAP
Monte Carlo simulation (shaded histogram): (a) \qe, (b) Bjorken-$x$, \xes\ and 
(c) the total hadronic center-of-mass energy,
\wes. The remaining plots show a
comparison of the reconstructed  \dspm-related quantities at the detector
level for data and for the RAPGAP 
Monte Carlo simulation: (d) \ptds, (e) \etads\ and (f) \xd.}
\label{f:mcdatadis}
\end{figure}

The properties of the selected events are 
compared with those of the RAPGAP Monte Carlo simulation.
All distributions 
shown are background-subtracted since they represent the number of signal 
events obtained by fitting the various mass distributions in a given 
bin. The data for the $K2\pi$ channel are shown in 
Figs.~\ref{f:mcdatacontrol} and \ref{f:mcdatadis} as solid 
points and the RAPGAP simulation as shaded histograms. All MC plots are 
normalized to have the same area as the data distributions.

Figure~\ref{f:mcdatacontrol} shows histograms of
\Ee, \te, \gh\ and \empz\ and
Fig.~\ref{f:mcdatadis}(a-c) displays the distributions of 
\qe, \xes\ and \wes. In general, reasonable agreement is observed 
between data and the MC simulation.
Figure~\ref{f:mcdatadis}(d-f) shows the transverse 
momentum, \ptds,  the pseudorapidity, \etads, and the
energy fraction carried by the \dspm\ in the $\gamma^*p$ center-of-mass frame,
\xd. Although the \ptds\ spectrum of the data is well described, the 
MC pseudorapidity spectrum is shifted to lower $\eta$ 
compared to the data and the \xd\ spectrum for the MC is shifted 
to slightly larger values. These discrepancies are 
examined in more detail below.

The HERWIG~\cite{herwig} Monte Carlo was 
used for systematic studies. This MC describes the \dspm\ 
differential distributions better
than RAPGAP, but does not give as good agreement
with the DIS variables as RAPGAP
since it does not contain QED radiative effects.

Photoproduction, where the final positron is scattered through very small
angles and escapes undetected through the RCAL beamhole, is a possible 
background source. Hadronic activity in
the RCAL can be wrongly identified as the scattered positron, giving rise to
fake DIS events. The effect was investigated using a large
sample of photoproduced \dspm\ events generated with the HERWIG MC. 
After the final selection cuts the photoproduction contamination 
is found to be less than 1\%, much smaller than the statistical error of the
measurement, and so is neglected.

The overall contribution to \dspm\ production from $b$ quark decays 
in the measured kinematic
region is estimated to be less than 2\%,
using HVQDIS with $m_b = 4.5$ GeV 
and a hadronization fraction of $b \rightarrow D^{\ast}$ of 
0.173~\cite{OPAL222}. A similar study using
RAPGAP yields an estimate of $\sim $1\% at low $Q^2$ and less than 
3\% at high $Q^2$. Hence the contribution from $b$ quark decays has been
neglected.

\section{Systematic uncertainties}
The experimental systematic uncertainties in the cross section are grouped into several 
major categories:
\begin{itemize}
\item
Systematic uncertainties related to the inclusive DIS selection of the 
events: variations were made in the $y$ cut, the RCAL box cut,
and the vertex position cut. In addition, for the $K2\pi$ final state
both $Q^2$ and $y$ were determined
using the $DA$ method rather than using the 
$e\Sigma$ method. 
The combined variations resulted in a change of $\pm$1.5\% to the nominal 
cross section. 
For the $K4\pi$ channel,
the electron method was used rather than the $DA$ method and
the combined variations resulted in a change of $\pm 4.7$\%.

\item 
Systematic uncertainties in the \dspm\ selection:
for the $K2\pi$ final state, the minimum transverse momentum of tracks used 
in the \dspm\ reconstruction 
was raised and lowered by up to 100 MeV for the $K$ and $\pi$ and by
25 MeV for the $\pi_s$ and resulted in a $\pm$4.5\% variation. The
momentum ratio $p(\dz)/p(\ps) $ was raised by $+0.5$ and yielded
a negligible change. For the
$K4\pi$ channel, similar changes in the minimum transverse momenta
and varying $p(\dz)/p(\ps) $ by $\pm 0.5$ combine to 
give a $\pm 8.5$\% variation.

\item
Systematic uncertainties related to the estimation of the number of events and
background uncertainty: for the $K2\pi$ analysis, a $\chi^2$ \dm-fit 
instead of a (binned) logarithmic-likelihood fit was performed. 
The \mdz\ signal range, within which the \dm\ distributions were fitted,
was varied by $\pm 10$~MeV. 
These variations resulted in a change of $\pm$2.5\%.
 For the $K4\pi$ analysis, the \mdz\ signal range
was also varied by $\pm 10$~MeV
and the width of the side-bands used to estimate combinatorial background
was varied. These variations resulted in a change of $\pm 5.8$\%. 

\item
The systematic uncertainty related to the MC generator was estimated 
for the $K2\pi$ analysis by using the HERWIG MC 
generator to calculate the corrections for the cross section 
determination. This variation yields a change of -1.3\%.
The larger overall systematic uncertainty for the $K4\pi$ channel means that 
this systematic uncertainty was negligible for this channel.

\item 
Since approximately 10\% of the \dspm\  events~\cite{vancouver}
are produced through a diffractive mechanism
 which was not included in the Monte Carlo generators
used to correct the data, acceptance corrections have also been obtained
using a sample of diffractive events generated with
RAPGAP. The difference in the global 
correction factor for the diffractive events was less than 10\%. This yielded
a 1.3\% variation in the overall cross section and was neglected.

\item
The systematic uncertainty related to the trigger for the $K2\pi$ final state
was estimated to be $\pm$1\%
by an analysis using only the inclusive DIS triggers  and was neglected.

\item 
The overall normalization uncertainties due to the luminosity measurement
error of $\pm$1.65\%,  and those due to the \dspm\ and $D^0$ decay branching 
ratios~\cite{PDG}  were not included in the systematic uncertainties.
\end{itemize}

The systematic uncertainties were added in quadrature. The total systematic 
uncertainty is less than 
the statistical error for most of the differential distributions and is
of the same order as the statistical error for the integrated cross section.

\section{Cross sections}

The cross sections for a given observable $Y$ were determined from the equation:
\begin{equation}
\frac {d\sigma}{dY} = \frac {N } {A \cdot \mathcal {L} \cdot B \cdot
\Delta Y}
\end{equation}
where $N$ is the number of \dspm\  events 
in a bin of $\Delta Y$, $A$ is the acceptance (including migrations,
efficiencies and radiative effects)
for that bin, $\mathcal {L}$ is the integrated luminosity and $B$ is
the product of the appropriate branching ratios for the \dspm\ and 
\dz.

The RAPGAP MC was used to estimate the acceptance. In the $K2\pi$ 
($K4\pi$) kinematic region the overall acceptance was 25.4 (18.0)\%. 
The statistical error of the MC is negligible compared to that of the data. 

The measured  cross section in the region $1<Q^2<600$ GeV$^2$, $0.02<y<0.7$,
for $1.5<\ptds<15$ GeV and $|\etads|<1.5$ using the $K2\pi$ final state, is 
\begin{center}
$\sigma(e^+p \rightarrow e^+D^{*\pm}X) = 
8.31 \pm 0.31(\mbox{stat.})^{+0.30}_{-0.50}(\mbox{syst.}) \mbox{ nb}$, 
\end{center}
and for $2.5<\ptds<15$ GeV and $|\etads|<1.5$ the cross section 
using the $K4\pi$ channel is 
\begin{center}
$\sigma(e^+p \rightarrow e^+D^{*\pm}X) = 
3.65 \pm 0.36(\mbox{stat.})^{+0.20}_{-0.41}(\mbox{syst.}) \mbox{ nb}$. 
\end{center}
These results are in good
agreement with the 
HVQDIS~\cite{hvqdis} calculations of 8.44 and 4.13~nb, respectively.
The parton
distribution functions resulting
from a ZEUS NLO QCD fit~\cite{zeus-nlo} 
to the ZEUS, NMC and BCDMS data were used in these calculations. In this
fit,
only the gluon and three light quark flavors were assumed to be present.
HVQDIS fragments the charm quark to a 
\dspm\ using the Peterson fragmentation function.
For the above calculation, \mc\ was set to
1.4~GeV, the Peterson fragmentation parameter $\epsilon$=0.035, and the mass 
factorization and renormalization scales were both set to $\sqrt{4m_c^2 
+\qsq}$. The hadronization
fraction $f(c\rightarrow\dsp)$ was set to $0.222\pm 0.014 \pm 0.014$, as 
measured by the OPAL collaboration~\cite{OPAL222}.  The result is, 
however, sensitive to the choice of the parameters. 
For example, varying \mc\  from 1.3 to 1.5~GeV
results in a variation of $\pm$0.55~nb to the $K2\pi$ prediction of
8.44 nb. If, instead
of using the ZEUS NLO parton distributions, the three-flavor
GRV98HO~\cite{grv98} set is used, the calculated cross section
for the $K2\pi$ kinematic region, assuming a charm mass of 1.4 GeV, 
 is reduced from 8.44 nb to 7.00~nb.

\begin{figure}[hbtp]
\epsfysize=15cm
\centerline{\epsffile{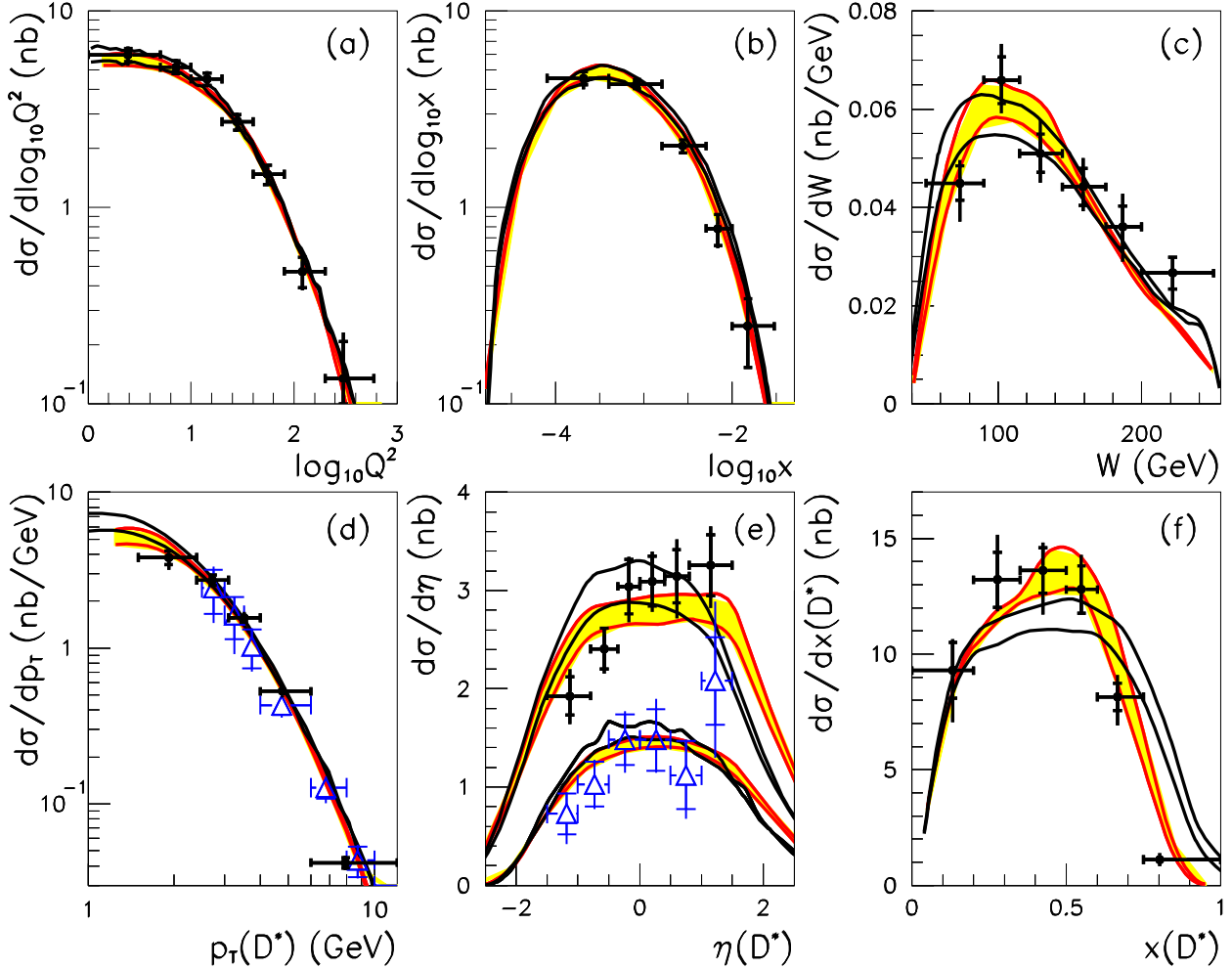}}
\caption{\it Differential cross sections for \dspm\ production
from the $K2\pi$ final state (solid dots) in the \qsq, $y$, 
\ptds\ and \etads\ kinematic region as functions of (a) $\log_{10}\qsq$, (b) 
$\log_{10}x$, (c) $W$, (d) \ptds, (e) \etads\ and (f) \xd. The inner 
error bars show the statistical uncertainties while the outer ones show the 
statistical and systematic uncertainties summed in quadrature. The results from 
the $K4\pi$ channel (open triangles) are also shown in the \ptds (d) and 
\etads\ (e) plots. 
The data are compared with the NLO QCD calculation as implemented in 
HVQDIS using the ZEUS NLO pdf's. The open band corresponds to the
standard Peterson fragmentation function with the parameter
$\epsilon$ = 0.035. For the shaded band,
the Peterson fragmentation was replaced by that extracted from
RAPGAP (see the text for details).
The boundaries of the bands correspond to charm mass 
variations between 1.3 (upper curve) and 1.5 \GeV (lower curve).
In (a) and (b), the open band is indistinguishable from the 
shaded band.
}
\label{f:diffxsections}
\end{figure}

\begin{figure}[hbtp]
\epsfysize=15cm
\centerline{\epsffile{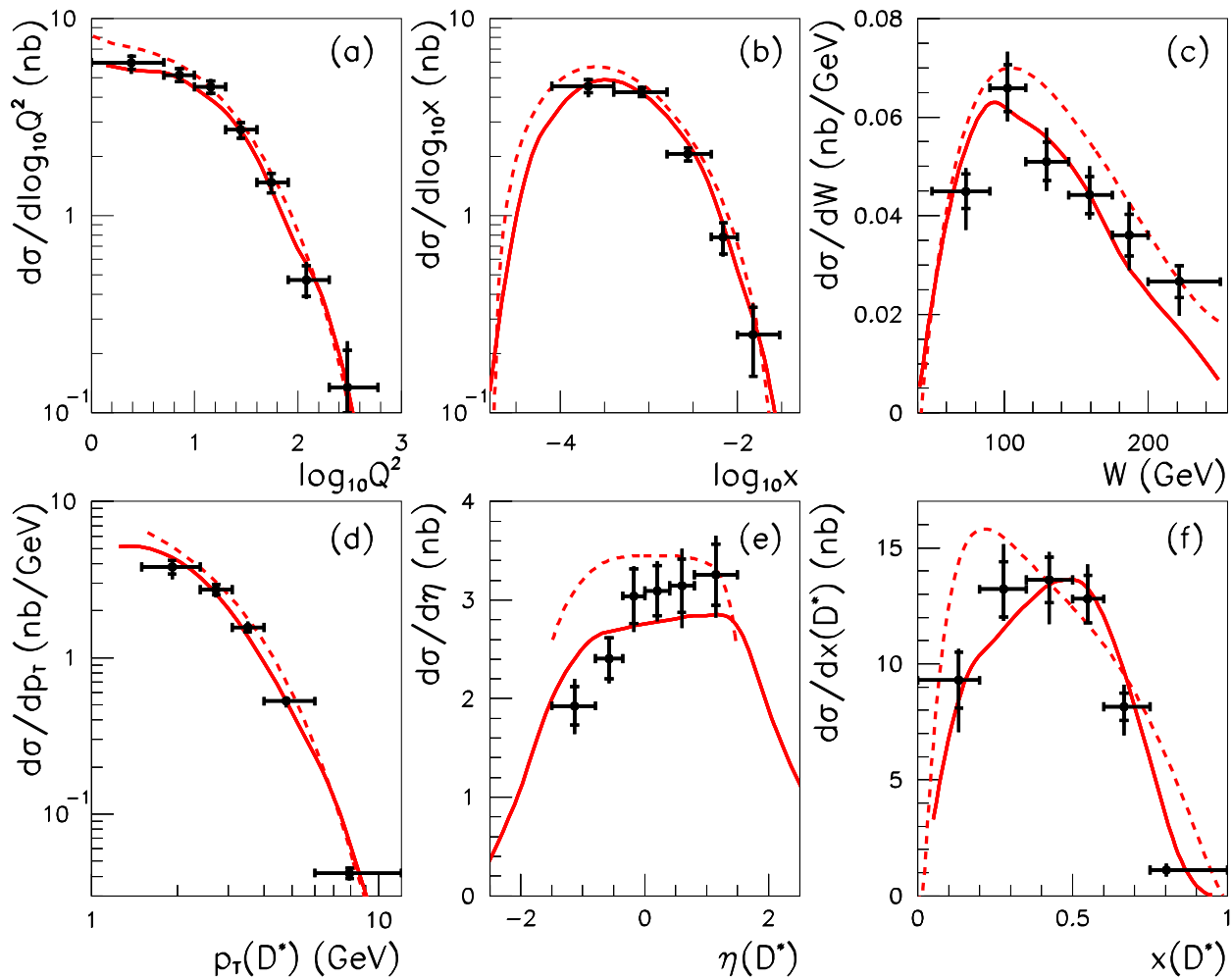}}
\caption{\it Differential cross sections for \dspm\ production
from the $K2\pi$ final state (solid dots) in the accessible \qsq, $y$, 
\ptds\ and \etads\ region as functions of (a) $\log_{10}\qsq$, (b) 
$\log_{10}x$, (c) $W$, (d) \ptds, (e) \etads\ and (f) \xd. The inner 
error bars show the statistical uncertainties while the outer ones show the 
statistical and systematic uncertainties summed in quadrature. 
The solid curves show the results of the HVQDIS calculation with 
RAPGAP-based fragmentation and $m_c$ = 1.4 \GeV while the dashed curves
correspond to the BKL results (see text).}
\label{f:ir1}
\end{figure}

Figures~\ref{f:diffxsections} and \ref{f:ir1} show 
the \dspm\ differential cross sections 
in the measured kinematic regions. The data points are drawn at the bin 
center of gravity, which is defined as the point at which the value of the 
assumed theoretical curve (HVQDIS with central choice of parameters and
RAPGAP fragmentation, see the discussion below) 
equals the mean value of the curve in the bin.
The cross sections are compiled in
Table~\ref{t:kp1}.
The 9\% error on $f(c\rightarrow\dsp)$ introduces an 
extra normalization uncertainty on the theoretical predictions 
which is not shown. 

The data are compared with two different theoretical calculations.
There is good agreement with HVQDIS, shown as the open bands
in Fig.~\ref{f:diffxsections}, 
except for the \etads\ and \xd\ distributions, where 
the HVQDIS calculations show a shift with respect to the data
to more negative \etads ~values and larger \xd\ values.


The disagreement between the \etads ~distribution and HVQDIS 
has also been observed by the H1 collaboration in
their DIS charm production data~\cite{h195}. 
A large discrepancy
in the \etads\ distribution is also observed~\cite{zeus_ds_php} between
the ZEUS photoproduction of charm and a massive NLO calculation~\cite{frixione}.
The discrepancy in shapes between the data and the HVQDIS prediction
could result from the use of the Peterson
fragmentation function. This predicts
the magnitude of the momentum of the $D^{\ast +}$ from
$c \rightarrow D^{\ast +}$ but produces 
no $D^{\ast +}$ transverse momentum relative to the $c$ quark.
In addition, no QCD evolution is included.
In contrast, the 
fragmentation models in JETSET and HERWIG predict a migration
of the charm quark emerging from the hard interaction towards positive
pseudorapidities as it fragments into a $D^{\ast +}$ due to
the interaction between the color charges of the $c$ quark and the
proton remnant. This has been called
the `beam-drag effect'~\cite{norrbin}.

To quantify the contribution of these fragmentation effects to the predicted
\dspm\ differential cross sections, the RAPGAP Monte Carlo charm quark
distribution was 
reweighted to match the two dimensional $p_T(c), ~\eta(c)$ distributions for
the charm quarks from HVQDIS. The fragmentation of these quarks 
was simulated by the Monte Carlo program
including parton shower effects\footnote{Note that this procedure 
keeps the cross
section in the overall phase space in \ptds\ and \etads ~equal to the HVQDIS 
result and is equivalent to applying the RAPGAP fragmentation to the HVQDIS 
charm quark prediction.}.
This reweighting procedure leads to an 
improved description of the data, as can be seen in Fig.~\ref{f:diffxsections}
(shaded bands),
where the two different implementations of the $c \rightarrow D^{\ast +}$
fragmentation are compared. 
In order to study systematic effects, the
same procedure as above was followed but with
the parton shower option switched off and HERWIG used. This produced 
qualitatively similar
results to the nominal method\footnote{
Some double counting may occur between the NLO calculations and the parton 
shower, which contains resummed terms of all orders. This effect is estimated
to be small from the calculations with the parton shower option in
JETSET switched off.}. The data are again compared to the HVQDIS calculation
including the RAPGAP-based fragmentation in Fig.~\ref{f:ir1}. The solid
curves are from using $m_c = 1.4$ GeV, corresponding to the central
mass value of the shaded band in Fig.~\ref{f:diffxsections}.

The data are also compared in Fig.~\ref{f:ir1} 
with the BKL tree-level calculations~\cite{bkl} 
(dashed curves) in which the \dspm\ is created from both color 
singlet and color octet contributions to the $(c\bar{q})$-state.
The relative weight of these color configurations
has been taken from comparisons of the BKL calculation to the published ZEUS 
\ds\ cross sections in photoproduction~\cite{zeus_ds_php}.
The calculation shows reasonable agreement with these DIS data.

Tables~\ref{t:kp} and \ref{t:f2c} give the resulting integrated cross sections
binned in $Q^2$ and $y$ for
the $K2\pi$ and $K4\pi$ final states, respectively.
The bin 
widths were chosen such that they contained of the order of 
100 signal events. The resulting purity in each bin is better than 70\%. The
resolutions in both $Q^2$ and $y$ are better than 10\% in all bins.
These bins were chosen to measure \ftwoccb\ as described in the next section.
Tables~\ref{t:kp} and \ref{t:f2c} also show
the HVQDIS predictions for the different kinematic bins. The
predictions are in good agreement with the data.

The quantitative agreement obtained between
data and the HVQDIS calculation displayed in Fig.~\ref{f:diffxsections}
and Tables~\ref{t:kp} and \ref{t:f2c} 
represents a confirmation of
the hard scattering factorization theorem, in that the same gluon and three
light-quark-flavor parton distributions describe both the ZEUS $F_2$ data
and the \dspm\ differential cross sections reported here.
In view of the agreement observed here, the HVQDIS program can be used to 
extrapolate outside the accessible kinematic region to obtain the
total \dspm\ cross section.

\section{Extraction of {\bf \ftwoccb}}
\label{s:extraction}

The charm contribution, \ftwoccb, to the proton structure function 
\ftwo\ can be related to the double differential $c\bar{c}$ cross section in 
$x$ and \qsq\ by
\begin{equation}
\frac{d^2\sigma^{c\bar{c}} (x, Q^2)}{dxdQ^2} = 
\frac{2\pi\alpha^2}{x Q^4}
\{ [1+(1-y)^2] F_2^{c\bar{c}}(x, Q^2) - y^2 F_L^{c\bar{c}}(x, Q^2) \} .
\end{equation}
In this paper, the $c\bar{c}$ cross section is obtained by measuring the
\dspm\ production cross section and employing the hadronization fraction
$f(c \rightarrow D^{\ast +})$ to derive the total charm cross section.
Since only a limited kinematic region is accessible for the measurement
of \dspm, a prescription for extrapolating to the full kinematic phase
space is needed. The contribution of $F_L^{c\bar{c}}(x, Q^2)
$ to the cross section in the measured 
\qsq, $y$ region is estimated from the NLO theoretical 
prediction~\cite{riemersma} to be less than 1\%
and is therefore neglected.
Equation (4) defines \ftwoccb\ as 
arising from events with one or more charm particles in the
final state, but it is not a unique theoretical definition.
It depends on the scheme
and parton distributions~\cite{CTEQpreprint}.

In order to measure the contribution of charm to
the inclusive  \ftwo, the integrated cross sections 
in the $Q^2$ and $y$ kinematic bins of Tables~\ref{t:kp} and \ref{t:f2c} 
were extrapolated to the full \ptds\ and \etads\ phase space using HVQDIS
with the RAPGAP-based fragmentation corrections discussed 
in the previous section.
Typical extrapolation factors for the $K2\pi$ ($K4\pi$) final state
were between 4 (10), at low \qsq, 
and 1.5 (4), at high \qsq.
This procedure neglects the possibility of additional
contributions outside the measured region due, for example, to 
intrinsic charm~\cite{golub}.

The extrapolated cross sections are converted into  $c\bar{c}$ 
cross sections using the hadronization fraction of 
charm to \dsp: $f(c\rightarrow\dsp) = 0.222\pm 0.014\pm 0.014$~\cite{OPAL222}.
The use of this value from OPAL implicitly assumes that 
charm production in DIS and $e^+e^-$ annihilation produces the same fractions
of the various
charm final-states. The production of charm bound states, such as 
$e^+p\rightarrow e^+J/\psi X$, is not accounted 
for when using the LEP $c\rightarrow \dsp$ branching fraction.
However, the inelastic $J/\psi$ cross section has been calculated~\cite{jpsi} to be only
2.5-4.5\% of the total charm production cross section predicted by HVQDIS 
in the $Q^2$, $y$ range of
this analysis. The elastic $J/\psi$ cross section has been measured in 
DIS~\cite{jpsiel} and is less than 0.5\% of the predicted total charm cross
section in the range of that measurement. The 9\% uncertainty on the $c \rightarrow D^{\ast +}$ hadronization is larger than that
arising from these $J/\psi$ contributions, which have consequently been neglected.

The systematic uncertainty on the extrapolation of the measured \dspm\ cross 
sections to the full \ptds\ and \etads\ phase space 
was investigated: varying the parameter $m_c$ by $\pm 0.15$ GeV gave a
variation which was typically $<$5\%;
using the standard Peterson 
fragmentation parameter $\epsilon = 0.035$ (instead of the RAPGAP fragmentation
correction) yielded changes typically $<$15\%; and using the GRV98HO pdf's 
in the NLO calculation generally caused changes of $<$20\%. 
If these uncertainties are added
in quadrature, they are typically smaller than the statistical errors.
However, the fact that the data are measured
in a small part of the available phase space means that a realistic
uncertainty on the extrapolation cannot be evaluated.
Therefore these extrapolation uncertainties are not included in the systematic
uncertainties discussed below. 

Since the structure 
function varies only slowly,  it is assumed to be constant within a given
\qsq\ and $y$ bin, so that the measured \ftwoccb\ 
in a bin $i$ is given by 
\begin{equation}
\ftwoccb_{meas}(x_i, Q^2_i) = 
                   \frac{\sigma_{i,meas}(e^+p\rightarrow \ds X)}
                        {\sigma_{i,theor}(e^+p\rightarrow \ds X)}
                    \ftwoccb_{theor}(x_i, Q^2_i)
\end{equation}
where  the cross sections $\sigma_i$ in bin $i$ are those for the measured  
\ptds\ and \etads\ region, and the subscripts
$meas$ and $theor$ denote `measured' and `theoretical', respectively.
The value of $\ftwoccb_{theor}$ was calculated from the NLO coefficient 
functions~\cite{smithlaenenneerven}, 
as implemented in a convenient parametrization~\cite{riemersma}. 
The functional form of $\ftwoccb_{theor}$ 
was used to quote the results for \ftwoccb\
at convenient values of 
$x_i$ and $ Q^2_i$ close to the center-of-gravity of the bin.
In this calculation, the same 
parton densities, charm mass (\mc = 1.4 GeV), and 
factorization and renormalization scales ($\sqrt{4m_c^2 + \qsq}$)
have been used as for 
the HVQDIS calculation of the differential cross sections. 

\subsection{Combination of \ftwoccb\ from both decays}

Finally, the results from the two decay channels were combined in the eight
common bins, taking into 
account all systematic uncertainties. 
The combined value is a weighted average with weights according to the 
statistical precision of the individual measurements. 
The systematic
uncertainties were assumed to be either uncorrelated or 100\% correlated 
between the analyses, as appropriate. 
All uncertainties concerning the DIS event
selection were assumed to be correlated. 
Only the effect of the variation of the \dz\ mass window and
the changes in the $p_T$ requirements for the \dz\ decay products 
were taken as uncorrelated.
As in both analyses, the positive and negative errors were treated separately.
The procedure leads to a gain in 
statistical precision of 5-25\%, compared to using only the $K2\pi$ decay
channel.

\subsection{Results and discussion}
\label{s:results}
\begin{figure}[hbtp]
\hspace{2.5cm}
\epsfysize=18cm
\centerline{\epsffile{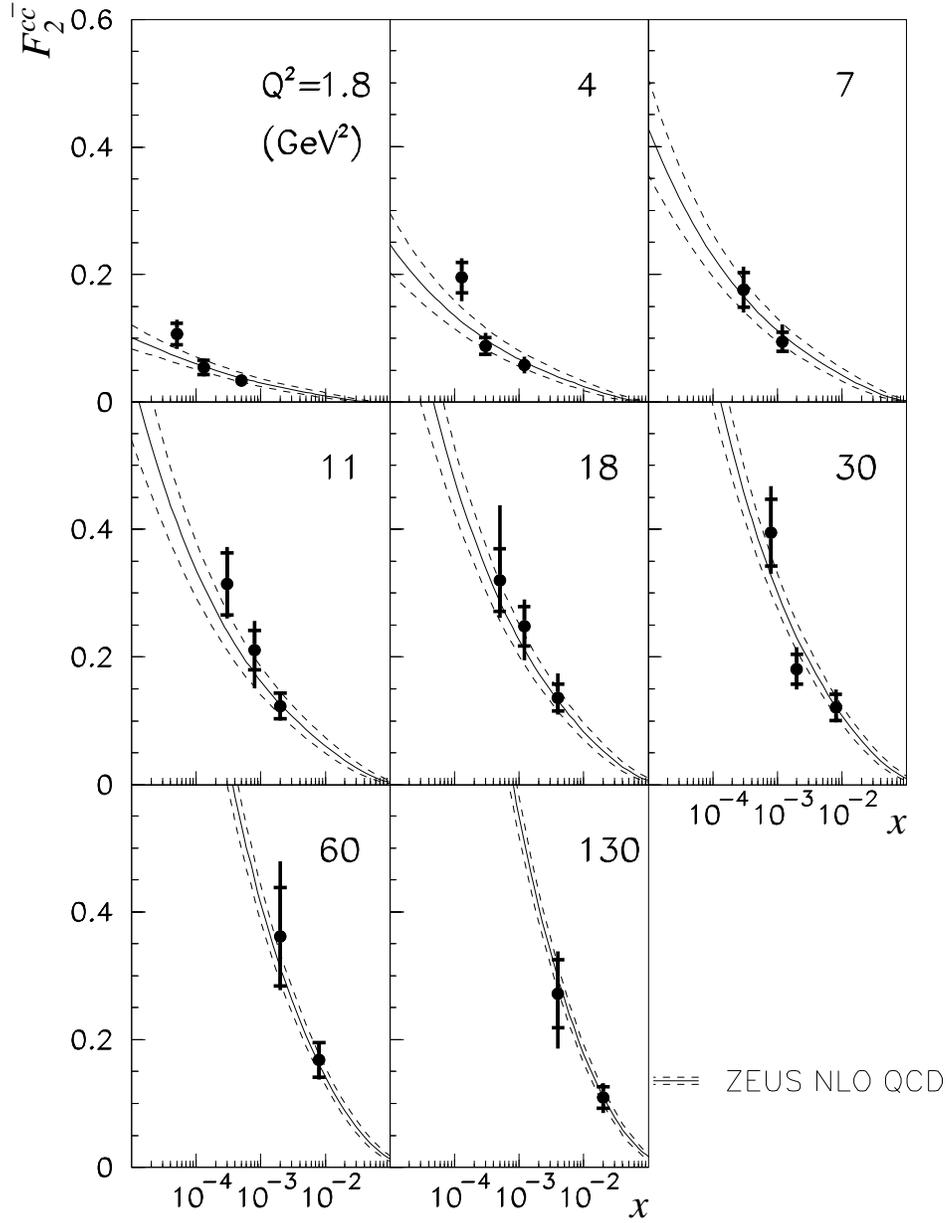}}
\caption{\it The measured \ftwoccb\ at \qsq\ values between 1.8 and 130 \GeVs
as a function of $x$. The inner error bars show the statistical uncertainty
and the outer ones show the statistical and experimental
systematic uncertainties summed in quadrature. 
The curves correspond to the NLO QCD calculation~\protect\cite{smith,riemersma}
 using the result of the ZEUS NLO 
QCD fit to \ftwo~\protect\cite{zeus-nlo}. The solid curves correspond
to the central values and the dashed curves give the uncertainty due 
to the parton distributions from the ZEUS NLO fit.
The overall normalization uncertainties arising from 
the luminosity measurement ($\pm $1.65\%), the \dspm\ and $D^0$ 
decay branching ratios, the charm hadronization fraction to $D^{\ast +}$
 ($\pm 9$\%) and the extrapolation
uncertainties (see text) are not included.}
\label{f:f2cvsx}
\end{figure}

Table~\ref{t:kp2} and  Fig.~\ref{f:f2cvsx} display the 
\ftwoccb\ values in the various \qsq\ bins 
as a function of $x$. 
The structure function \ftwoccb\ shows a rise with decreasing $x$ at 
constant values of \qsq. The rise becomes steeper at higher \qsq.

The curves in Fig.~\ref{f:f2cvsx}
represent the results of the NLO QCD 
calculation~\cite{riemersma}
with the ZEUS NLO QCD pdf's. The central, solid
curve corresponds to a charm quark mass  of 1.4~GeV.
Since good agreement was obtained 
between data and the HVQDIS calculation for the \dspm\ differential 
cross sections and for the integrated cross sections shown in
Tables~\ref{t:kp} and \ref{t:f2c} 
and since that calculation was used to extrapolate to the
full kinematic range, the curves would be expected to describe the resulting 
values of \ftwoccb. The total uncertainty in the calculation of \ftwoccb,
shown as the band of dashed curves around the solid curve, corresponds
to the uncertainty propagated from the ZEUS NLO QCD fit and
is dominated by the uncertainty in the charm quark mass, which was
varied from 1.2 to 1.6 GeV.

\begin{figure}[hbtp]
\epsfysize=18cm
\centerline{\epsffile{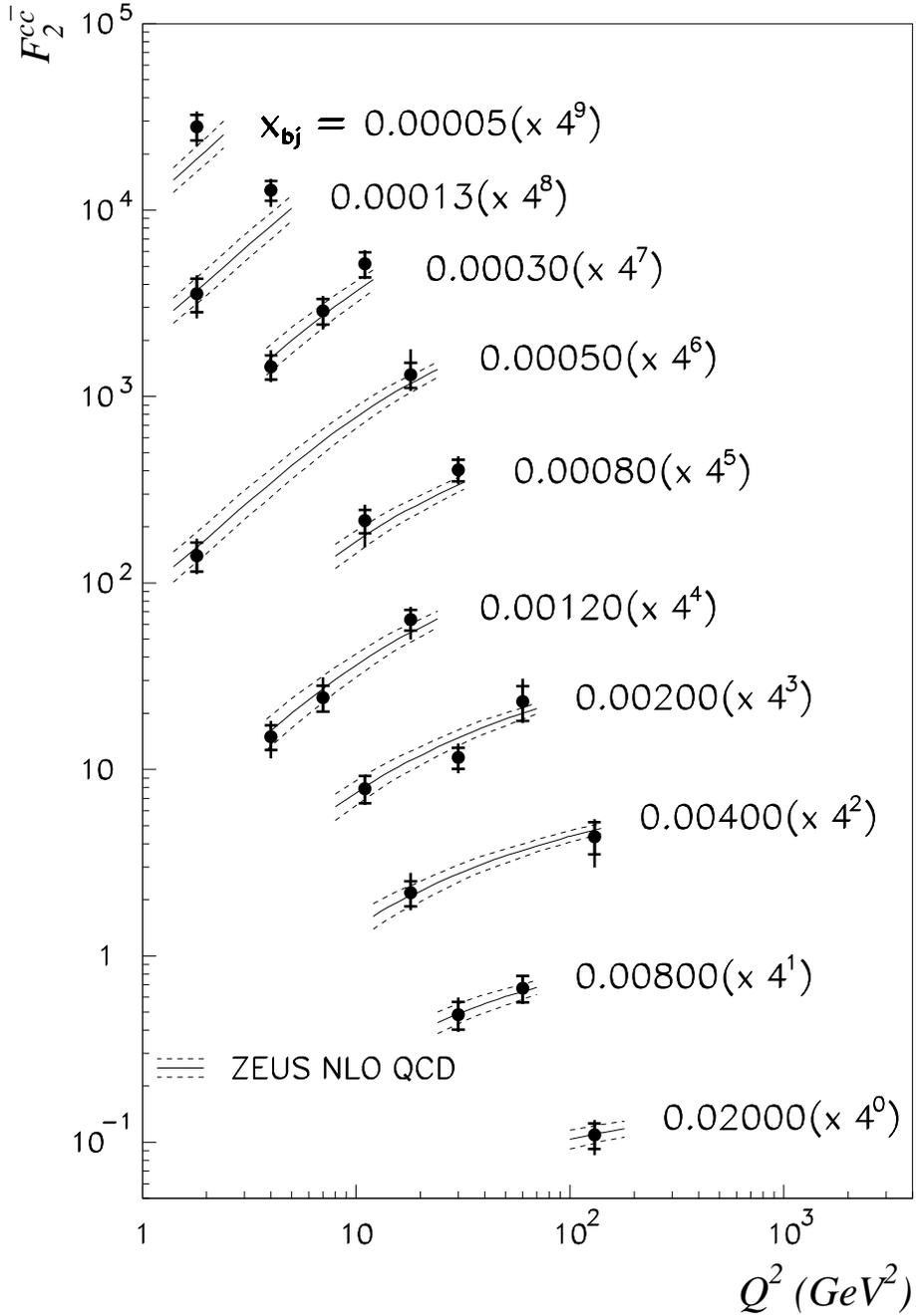}}
\caption{\it The measured \ftwoccb\ at $x$ values between 0.00005 and 0.02 
as a function of \qsq. The various values of $x$ are indicated to the 
right of the data points. For clarity of presentation, the
\ftwoccb\ values have been 
scaled by the number shown in parentheses next to the $x$ value. 
The inner error bars show the 
statistical uncertainty and the outer ones show the statistical
and systematic uncertainties summed in quadrature. The curves
correspond to the 
NLO QCD calculation~\protect\cite{smith,riemersma} using
the result of the ZEUS NLO QCD fit to \ftwo~\protect\cite{zeus-nlo}.
The solid curves correspond
to the central values and the dashed curves give the uncertainty due 
to the parton distributions from the ZEUS NLO fit.
Details of this calculation are given in the text.
The overall normalization uncertainties arising from 
the luminosity measurement ($\pm $1.65\%), the \dspm\ and $D^0$ 
decay branching ratios, the charm hadronization fraction to $D^{\ast +}$
 ($\pm 9$\%) and the extrapolation
uncertainties (see text) are not included.}
\label{f:f2cvsq2}
\end{figure}

Figure~\ref{f:f2cvsq2} shows \ftwoccb\ at constant $x$ values as a
function of \qsq. Although the number of points is small, large scaling
violations of the structure function are evident. 
The curves superimposed on the data are from the same calculation as 
shown in Fig.~\ref{f:f2cvsx}.

\begin{figure}[hbtp]
\hspace{2.5cm}
\epsfysize=18cm
\centerline{\epsffile{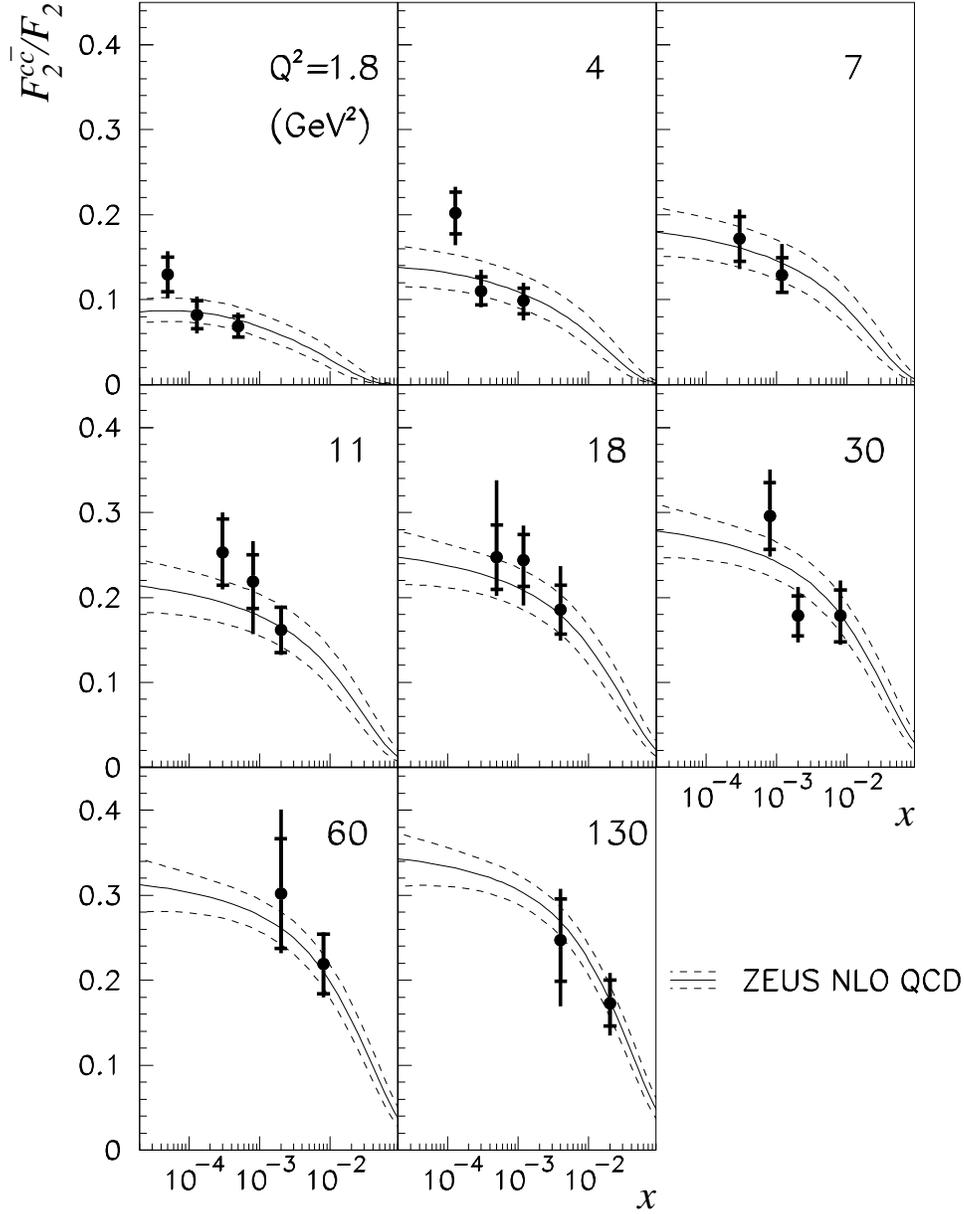}}
\caption{\it The ratio of \ftwoccb\ to \ftwo\ at \qsq\ values between 1.8 and 
130 \GeVs as a function of $x$.  The ratio is calculated using the 
\ftwoccb\ values measured in this analysis and the ZEUS NLO QCD fit to 
\ftwo~\protect\cite{z95f2}. The inner 
error bars show the statistical uncertainty and the outer ones show the 
statistical and systematic uncertainties summed in quadrature. The curves
correspond to the NLO QCD calculation~\protect\cite{smith,riemersma} 
using the result of the ZEUS NLO QCD fit to \ftwo~\protect\cite{zeus-nlo}. 
The solid curves correspond
to the central values and the dashed curves give the uncertainty due 
to the parton distributions from the ZEUS NLO fit.
Details of this calculation are given in the text.
The overall normalization uncertainties arising from 
the luminosity measurement ($\pm $1.65\%), the \dspm\ and $D^0$ 
decay branching ratios, the charm hadronization fraction to $D^{\ast +}$
 ($\pm 9$\%) and the extrapolation
uncertainties (see text) are not included.}
\label{f:f2covf2}
\end{figure}

Figure~\ref{f:f2covf2} shows the ratio of \ftwoccb\  to \ftwo,
the inclusive proton
structure function, as a function of $x$ in fixed-$Q^2$ bins.
The curves superimposed on the data are again from the calculation used
for Fig.~\ref{f:f2cvsx}.
The values of \ftwo\ used to determine the ratio were
taken from the ZEUS NLO QCD fit at the same $Q^2$ and  $x$ 
for which
\ftwoccb\ is quoted. The error on \ftwo\ is negligible in comparison
to \ftwoccb. The charm 
contribution to \ftwo\  rises steeply with decreasing $x$.
In the measured $x$ region, \ftwoccb\ accounts for $<10$\% of \ftwo\ at 
low \qsq\ and $x\simeq 5\cdot10^{-4}$ and rises to 
$\simeq 30$\% of \ftwo\ for \qsq$>11$~GeV$^2$ 
at the lowest $x$ measured. 
The strong rise of \ftwoccb\ at low values of $x$ is similar to that
of the gluon density and thus supports the hypothesis that charm production is
dominated by the boson-gluon-fusion mechanism.  
 
\section{Summary}
This paper presents an analysis of \dspm\ production
in DIS using the combined ZEUS 1996 and 1997 
data samples with an integrated luminosity of 37 pb$^{-1}$,
about ten times larger than in
the previous ZEUS 
study. In addition, both the $K2\pi$ and $K4\pi$
decay modes of the $D^{\ast}$ have been employed and their results combined. 
In the experimentally 
accessible region of 1.5 (2.5) $ <$\ptds$<$15~GeV and $|\etads|<1.5$, the cross
section for \dspm\ production for the $K2\pi$ ($K4\pi$) final state 
in events with 1 $<$\qsq$<600$~GeV$^2$ and $0.02<y<0.7$ is 
$ 8.31 \pm 0.31(\mbox{stat})^{+0.30}_{-0.50}(\mbox{sys}) \mbox{ nb}$
($ 3.65 \pm 0.36(\mbox{stat})^{+0.20}_{-0.41}(\mbox{sys}) \mbox{ nb}$).

QCD calculations of charm production based on the NLO boson-gluon-fusion
process with three flavors of light quarks show 
excellent agreement with
the overall cross section and with the \qsq\ and $y$ distributions.
The \etads\ and \xd\ distributions, however, cannot be reproduced with the
standard Peterson fragmentation. Good agreement is obtained 
after a more appropriate $c \rightarrow D^{\ast +}$ fragmentation,
such as that in JETSET, is used.

The quantitative agreement between the NLO pQCD calculations and the ZEUS data
provides a confirmation of the hard scattering factorization
theorem, whereby the same gluon density in the proton 
describes both the inclusive $F_2$ and
the DIS production of charm.

The charm 
contribution, \ftwoccb, to the proton structure function \ftwo\ was obtained
using the NLO QCD calculation to extrapolate outside the measured
\ptds\ and \etads\ region.
Compared to the previous ZEUS
study, the kinematic range has been extended down to \qsq\ = 1.8~GeV$^2$ and
up to \qsq\ = 130~GeV$^2$, with reduced uncertainties.
 The structure function \ftwoccb\ exhibits 
large scaling violations, as well as a steep rise 
with decreasing $x$ at constant \qsq.
For \qsq $>$ 11 GeV$^2$ and $x \simeq 10^{-3}$, the 
ratio of \ftwoccb\ to \ftwo\ is about 0.3.

\section{Acknowledgements}
We thank the DESY Directorate for their strong support and
encouragement. The remarkable achievements of the HERA machine group
were essential for the successful completion of this work and are
gratefully appreciated.  We also acknowledge the many informative
discussions we have had with J. Amundson, A. Berezhnoy, J. Collins, 
S. Fleming, B. Harris, F. Olness,
C. Schmidt, J. Smith, W.K. Tung and A. Vogt.


\newpage

\clearpage

\begin{table}[hbpt]
\begin{center}
\caption{\it 
The \ds $\rightarrow K\pi \pi_s$ and \ds $\rightarrow K\pi \pi \pi \pi_s$
differential cross sections.
The bin range, the center-of-gravity of the bin 
(see text) and the
cross sections for all the data  in 
Figs.~\ref{f:diffxsections} and~\ref{f:ir1} are shown.
The first error is the statistical error and the asymmetric errors are the 
statistical and systematic uncertainties added in quadrature.
The overall normalization uncertainties arising from 
the luminosity measurement ($\pm $1.65\%) and from the \dspm\ and $D^0$ 
decay branching ratios are
not included.}
\vspace{0.5cm}
\begin{tabular}{| c | r @{$\pm$} l @{\extracolsep{0pt}} l || c | r @{ $\pm$ } l @{\extracolsep{0pt}}l |}
\hline
\multicolumn{4}{|c||}{$K\pi\pi_s$}&\multicolumn{4}{|c|}{$K\pi\pi_s$}\\
\hline
 (range) $log_{10}(\qsq)$ & \multicolumn{3}{c||}{$d\sigma/dlog_{10}(\qsq)$} & (range)  $log_{10}(x)$ & \multicolumn{3}{c|}{ $d\sigma/dlog_{10}(x)$}\\
 & \multicolumn{3}{c||}{(nb)} & & \multicolumn{3}{c|}{(nb)} \\
\hline 
       (0.0,0.7) 0.39 & 5.99  &  0.47  &$ ^{+  0.61 }_{-  0.74}$ & (-4.1,-3.4)  -3.69 &   4.54  &  0.35 &$ ^{+  0.45 }_{-  0.56}$ \\
     (0.7,1.0) 0.85 &  5.17   & 0.39 &$ ^{+   0.54 }_{-  0.46}$ & (-3.4,-2.8)  -3.08 &  4.24  &  0.23 &$ ^{+  0.34 }_{-  0.27}$ \\
     (1.0,1.3) 1.16 &  4.50  &  0.33 &$ ^{+  0.43 }_{-  0.44}$ & (-2.8,-2.3)  -2.56 &  2.06  &  0.16 &$ ^{+  0.19 }_{-  0.17}$ \\
     (1.3,1.6) 1.45 &  2.72  &  0.25 &$ ^{+  0.32 }_{-  0.29}$ & (-2.3,-2.0)  -2.16 &  0.78  &  0.14 &$ ^{+  0.15 }_{-  0.16}$ \\
     (1.6,1.9) 1.74 &  1.47  &  0.17 &$ ^{+  0.18 }_{-  0.17}$ & (-2.0,-1.5) -1.82 &  0.25  &  0.10 &$ ^{+  0.11  }_{- 0.10}$ \\
     (1.9,2.3) 2.08 &  0.47  &  0.08 &$ ^{+  0.11 }_{-  0.10}$ &  & \multicolumn{3}{c|}{} \\
     (2.3,2.8) 2.48 &  0.135  &  0.073 &$ ^{+  0.096 }_{-  0.073}$ &  &  \multicolumn{3}{c|}{}\\
\hline  \hline
\multicolumn{4}{|c||}{$K\pi\pi_s$}&\multicolumn{4}{|c|}{$K\pi\pi_s$}\\
\hline
 (range) $W$ & \multicolumn{3}{c||}{$d\sigma/dW$} & (range)  $\xd$ & \multicolumn{3}{c|}{ $d\sigma/d\xd$}\\
 (GeV)& \multicolumn{3}{c||}{(nb/GeV)} & & \multicolumn{3}{c|}{(nb)} \\
\hline 
    (50,90)  73 &  0.0450 & 0.0035 &$ ^{+ 0.0047}_{-  0.0079}$ & (0.0,0.2) 0.13  &   9.3  &    1.2 &$ ^{+   1.4 }_{-   2.2}$\\
(90,115) 102 &  0.0659&  0.0048 &$ ^{+ 0.0074}_{-  0.0068 }$& (0.2,0.35) 0.28  &  13.2  &   1.2 &$ ^{+   1.9 }_{-   1.4}$\\
(115,145) 129 &  0.0510&  0.0039&$ ^{+  0.0068}_{-  0.0059 }$& (0.35,0.5) 0.42 &   13.61  &   0.98  &$ ^{+   1.22 }_{-   1.90}$\\
(145,175) 159 & 0.0442 & 0.0038 &$ ^{+ 0.0059 }_{- 0.0049 }$& (0.5,0.6)  0.55  &  12.79   &  1.03 &$ ^{+    1.49  }_{-  1.03}$\\
(175,200) 187 & 0.0361&  0.0043 &$ ^{+ 0.0067}_{-  0.0071 } $& (0.6,0.75)  0.67  &   8.15   &  0.60  &$ ^{+   0.95  }_{-  1.24}$\\
(200,250) 222 &  0.0266&  0.0032 &$ ^{+ 0.0033 }_{- 0.0069 } $& (0.75,1.0)  0.80  &   1.12  &   0.10   &$ ^{+   0.27  }_{-  0.27}$\\

\hline  \hline
\multicolumn{4}{|c||}{$K\pi\pi_s$}&\multicolumn{4}{|c|}{$K\pi\pi\pi\pi_s$}\\
\hline
 (range) \ptds & \multicolumn{3}{c||}{$d\sigma/d\ptds$} & (range)  $\ptds$ & \multicolumn{3}{c|}{ $d\sigma/d\ptds$}\\
 (GeV)& \multicolumn{3}{c||}{(nb/GeV)} &(GeV) & \multicolumn{3}{c|}{(nb/GeV)} \\
\hline 
     (1.5,2.4)  1.91 & 3.82&  0.39&$ ^{+0.48}_{-0.65}$ &(2.5,3.0) 2.74  & 2.42&  0.76&$ ^{+  0.90 }_{- 0.97 }$  \\
     (2.4,3.1)  2.72 & 2.72&  0.21&$ ^{+  0.26 }_{-0.28}$&(3.0,3.5) 3.24 &1.63&  0.49&$ ^{+  0.52 }_{- 0.71 }$ \\
     (3.1,4.0)  3.50 & 1.57&  0.10&$ ^{+  0.14}_{- 0.13}$&(3.5,4.0) 3.73 & 1.03 & 0.29&$ ^{+ 0.31 }_{- 0.33 }$\\
     (4.0,6.0)  4.77 & 0.527& 0.033&$ ^{+  0.041}_{- 0.047}$ &(4.0,6.0) 4.74 & 0.43 & 0.06 &$ ^{+  0.07 }_{- 0.07 }$ \\
     (6.0,15.0)  7.93 & 0.0419& 0.0034&$ ^{+ 0.0043}_{- 0.0041}$ &(6.0,8.0) 6.77 & 0.127&  0.019&$ ^{+  0.022  }_{-  0.026 }$\\
 &  \multicolumn{3}{|c||}{} &(8.0,10.0) 8.75  &  0.044 & 0.010&$ ^{+  0.011 }_{- 0.010}$\\
 &  \multicolumn{3}{|c||}{} &(10.0,15.0) 11.73 & 0.007&  0.002&$ ^{+  0.003 }_{- 0.003 }$\\
\hline  \hline
\multicolumn{4}{|c||}{$K\pi\pi_s$}&\multicolumn{4}{|c|}{$K\pi\pi\pi\pi_s$}\\
\hline
 (range) \etads & \multicolumn{3}{c||}{$d\sigma/d\etads$} & (range)  $\etads$ & \multicolumn{3}{c|}{ $d\sigma/d\etads$}\\
 & \multicolumn{3}{c||}{(nb)} & & \multicolumn{3}{c|}{(nb)} \\
\hline 
    (-1.5,-0.8)  -1.13 &   1.93 &   0.19 &$ ^{+   0.27 }_{-   0.29} $  & (-1.5,-1.0) -1.18 & 0.73 &    0.21  &$ ^{+   0.22   }_{-  0.31}$\\
(-0.8,-0.35)   -0.58 &   2.41&    0.21 &$ ^{+   0.22 }_{-   0.25} $ & (-1.0,-0.5)  -0.73   & 1.03 &    0.23  &$ ^{+   0.25   }_{-  0.25}$\\
   (-0.35,0.0)  -0.18  &  3.04 &   0.28  &$ ^{+  0.30 }_{-   0.37}$ & (-0.5,0.0)   -0.24  & 1.48  &   0.26  &$ ^{+   0.29   }_{-  0.34}$ \\
    (0.0,0.4)   0.20 &   3.09 &   0.25  &$ ^{+  0.30}_{-    0.31}$ & (0.0,0.5)    0.26 & 1.48 &    0.31  &$ ^{+   0.42   }_{-  0.33}$ \\
   (0.4,0.8)   0.60  &  3.15 &   0.27  &$ ^{+  0.38 }_{-   0.43}$ & (0.5,1.0)    0.75 &  1.12  &   0.35  &$ ^{+   0.36  }_{-   0.44}$\\
    (0.8,1.5) 1.15 &   3.26  &  0.31 &$ ^{+   0.40 }_{-   0.44}$ &  (1.0,1.5)    1.22  & 2.08  &   0.45  &$ ^{+   0.81   }_{-  0.78}$\\
\hline

\end{tabular}
\label{t:kp1}
\end{center}
\end{table}
\newpage
\clearpage

\begin{table}[hbpt]
\begin{center}
\caption{\it The cross sections for $D^{\ast \pm}$ production from the $K2\pi$ 
final state. The table contains for each bin: the \qsq\ range of the bin;
the $y$ range of the bin;
the measured \dspm\ cross section in the bin
with statistical and systematic uncertainties; and the HVQDIS prediction for 
this cross section. 
The overall normalization uncertainties arising from 
the luminosity measurement ($\pm $1.65\%) and from the \dspm\ and $D^0$ 
decay branching ratios are
not included.}
\vspace{0.5cm}
\begin{tabular}{|c|c|c|c|}
\hline
 \qsq\ range    &   $y$       &$\sigma(\ds)$& $\sigma(\ds)$  \\ 
 (GeV$^2$)     &   range      &     meas. (nb) &     pred.(nb)\\ \hline
 1--3.5 &  0.70--0.24 & $1.45\pm 0.23^{+0.20}_{-0.22}$ & 1.12 \\ 
        &  0.24--0.11 & $1.00\pm 0.20^{+0.17}_{-0.17}$ & 1.00 \\ 
        &  0.11--0.02 & $0.92\pm 0.16^{+0.15}_{-0.09}$ & 1.10 \\ 
\hline
 3.5--6.5&  0.70--0.22 &$0.73\pm 0.10^{+ 0.07}_{-0.11}$ & 0.54 \\ 
         &  0.22--0.11 &$0.342\pm 0.055^{+ 0.068}_{-0.036}$ & 0.405 \\ 
         &  0.11--0.02 &$0.433\pm 0.066^{+ 0.069}_{-0.076}$ & 0.528 \\ 
\hline
 6.5--9  &  0.70--0.15 & $0.388\pm 0.060^{+0.051 }_{-0.053}$ &     0.369 \\ 
         &  0.15--0.02 & $0.288\pm 0.046^{+0.068}_{-0.008}$ & 0.355  \\ \hline

 9--14   &  0.70--0.23 &  $0.370\pm 0.057^{+0.038}_{-0.030}$ & 0.302 \\ 
         &  0.23--0.11 &$0.314\pm 0.045^{+0.051}_{-0.076}$ &     0.251 \\ 
         &  0.11--0.02 & $0.253\pm 0.042^{+0.011}_{-0.020}$ & 0.316
\\ \hline

 14--22  &  0.70--0.23 & $0.25\pm 0.05^{+0.11}_{-0.02}$ & 0.26\\ 
         &  0.23--0.11 &$0.254\pm 0.043^{+0.044}_{-0.073}$ &     0.212\\ 
         &  0.11--0.02 & $0.226\pm 0.035^{+0.052}_{-0.027}$ &     0.250 \\ \hline

 22--44  &  0.70--0.23 &$0.387\pm 0.053^{+0.052}_{-0.040}$ &     0.301 \\ 
         &  0.23--0.11 &$0.200\pm 0.027^{+0.028}_{-0.021}$ &     0.226\\ 
         &  0.11--0.02 &$0.198\pm 0.033^{+0.031}_{-0.018}$ & 0.240\\ \hline

 44--90  &  0.70--0.23 & $0.202\pm 0.043^{+0.050}_{-0.019}$ & 0.188 \\
        &  0.23--0.02 & $0.200\pm 0.032^{+0.011}_{-0.016}$ & 0.221\\ \hline
 90--200 &  0.70--0.23 & $0.090\pm 0.023^{+0.009}_{-0.013}$ & 0.099\\
         &  0.23--0.02 & $0.075\pm 0.015^{+0.010}_{-0.007}$ & 0.086\\ \hline
\end{tabular}
\label{t:kp}
\end{center}
\end{table}

\begin{table}[hbpt]
\begin{center}
\caption{\it The cross sections for $D^{\ast \pm}$ production from the $K4\pi$ 
final state. The table contains for each bin: the \qsq\ range of the bin; 
the $y$ range of the bin; 
the measured \dspm\ cross section in the bin
with statistical and systematic uncertainties; and the HVQDIS prediction for 
this cross section. 
The overall normalization uncertainties arising from 
the luminosity measurement ($\pm $1.65\%) and from the \dspm\ and $D^0$ 
decay branching ratios are
not included.}
\vspace{0.5cm}
\begin{tabular}{|c|c|c|c|}
\hline
 \qsq\ range     &   $y$       &$\sigma(\ds)$& $\sigma(\ds)$ \\
(GeV$^2$)        &  range      &     meas. (nb) & pred. (nb) \\ 
\hline
 1--10  & 0.70--0.34 & 0.94$\pm 0.29^{+0.39}_{-0.21} $  & 0.61\\
        & 0.34--0.02 & 1.39$\pm 0.51^{+0.21}_{-0.09} $  & 1.55\\
\hline
10--21  & 0.70--0.28 & 0.35$\pm 0.08^{+0.11}_{-0.06} $  & 0.21\\
        & 0.28--0.02 & 0.59$\pm 0.11^{+0.07}_{-0.11} $  & 0.47\\
\hline
21--33  & 0.70--0.22 & 0.115$\pm 0.055^{+0.039}_{-0.023} $     & 0.154\\
        & 0.22--0.02 & 0.125$\pm 0.045^{+0.048}_{-0.050} $     & 0.209\\
\hline
50--600 & 0.70--0.22& 0.23$\pm 0.07^{+0.07}_{-0.13} $     & 0.249\\
        & 0.22--0.02& 0.240$\pm 0.061^{+0.059}_{-0.084} $     & 0.221\\
\hline
\end{tabular}
\label{t:f2c}
\end{center}
\end{table}

\newpage
\clearpage

\begin{table}[hbpt]
\begin{center}
\caption{\it The \ftwoccb\ results derived from the combination 
of the $K2\pi$ and $K4\pi$ channels (see text).
The table contains for each bin: the \qsq\
value at which  \ftwoccb\ is reported; the $x$ value 
at which \ftwoccb\ is reported; and the 
measured \ftwoccb\ with statistical and systematic 
uncertainties. 
The overall normalization uncertainties arising from 
the luminosity measurement ($\pm $1.65\%), the \dspm\ and $D^0$ 
decay branching ratios, the charm hadronization fraction to $D^{\ast +}$
 ($\pm 9$\%) and the extrapolation
uncertainties (see text) are not included.}
\vspace{0.5cm}
\begin{tabular}{|c|c|c|}
\hline
\qsq\    &   $x$            &$F_{2}^{c\bar{c}}(x,Q^2)$      \\ 
(GeV$^2$)&                  & meas. $\pm $stat. $\pm$ syst. \\ \hline
1.8 & $5.0\cdot10^{-5}$ & $0.107\pm 0.017^{+0.015}_{-0.016}$ \\
    & $1.3\cdot10^{-4}$ & $0.054\pm 0.011^{+0.009}_{-0.009}$ \\ 
    & $5.0\cdot10^{-4}$ & $0.034\pm 0.006^{+0.005}_{-0.003}$ \\ \hline

4  & $1.3\cdot10^{-4}$ & $0.195\pm 0.024^{+0.018}_{-0.028}$ \\ 
   & $3.0\cdot10^{-4}$ & $0.088\pm 0.013^{+0.015}_{-0.008}$ \\ 
   & $1.2\cdot10^{-3}$ & $0.058\pm 0.009^{+0.009}_{-0.010}$ \\ \hline

7  & $3.0\cdot10^{-4}$ & $0.176\pm 0.027^{+0.023}_{-0.024}$ \\ 
   & $1.2\cdot10^{-3}$ & $0.095\pm 0.015^{+0.023}_{-0.002}$ \\ \hline

11 & $3.0\cdot10^{-4}$ & $0.314\pm 0.048^{+0.032}_{-0.025}$ \\ 
   & $8.0\cdot10^{-4}$ & $0.211\pm 0.030^{+0.035}_{-0.051}$ \\ 
   & $2.0\cdot10^{-3}$ & $0.123\pm 0.020^{+0.006}_{-0.010}$ \\ \hline

18 & $5.0\cdot10^{-4}$ & $0.32\pm 0.05^{+0.11}_{-0.03}$ \\ 
   & $1.2\cdot10^{-3}$ & $0.248\pm 0.031^{+0.028}_{-0.044}$ \\ 
   & $4.0\cdot10^{-3}$ & $0.136\pm 0.021^{+0.031}_{-0.017}$ \\ \hline

30 & $8.0\cdot10^{-4}$ & $0.395\pm 0.052^{+0.050}_{-0.037}$ \\ 
   & $2.0\cdot10^{-3}$ & $0.181\pm 0.024^{+0.025}_{-0.021}$ \\ 
   & $8.0\cdot10^{-3}$ & $0.121\pm 0.021^{+0.019}_{-0.011}$ \\ \hline

60 & $2.0\cdot10^{-3}$ & $0.361\pm 0.077^{+0.090}_{-0.033}$ \\ 
   & $8.0\cdot10^{-3}$ & $0.168\pm 0.027^{+0.009}_{-0.013}$ \\ \hline
130& $4.0\cdot10^{-3}$ & $0.272\pm 0.053^{+0.039}_{-0.067}$ \\ 
   & $2.0\cdot10^{-2}$ & $0.109\pm 0.017^{+0.014}_{-0.017}$ \\ \hline
\end{tabular}
\label{t:kp2}
\end{center}
\end{table}

%
%

%
%

%
%

%
%

%
%

%
%

%
%

%
%

%
%


\begin{thebibliography}{99}
\bibitem{zeus-nlo} The result of the ZEUS NLO QCD fit is described in
\cite{z95f2}. New fits were made using various charm masses
\end{thebibliography}
\end{document}